\documentclass[a4paper,11pt]{article}
\usepackage{jheppub}
\usepackage[T1]{fontenc}

\newcommand{\BE}{\begin{equation}}
\newcommand{\EE}{\end{equation}}
\newcommand{\BA}{\begin{eqnarray}}
\newcommand{\EA}{\end{eqnarray}}

\title{\boldmath Creation of D9-brane--anti-D9-brane Pairs from Hagedorn Transition of Closed Strings}

\author[a]{Kenji Hotta,}

\affiliation[a]{Department of Physics, Hokkaido University, Sapporo, Hokkaido, 060-0810, Japan}

\emailAdd{khotta@particle.sci.hokudai.ac.jp}

\abstract{It is well known that one-loop free energy of closed strings diverges above the Hagedorn temperature. One explanation for this divergence is that a `winding mode' in the Euclidean time direction becomes tachyonic above the Hagedorn temperature. The Hagedorn transition of closed strings has been proposed as a phase transition via condensation of this winding tachyon. However, we have not known the stable minimum of the potential of this winding tachyon so far. On the other hand, we have previously calculated the finite temperature effective potential of open strings on D-brane--anti-D-brane pairs, and shown that a phase transition occurs near the Hagedorn temperature and D9-brane--anti-D9-brane pairs become stable. In this paper, we present a conjecture that {\it D9-brane--anti-D9-brane pairs are created by the Hagedorn transition of closed strings}, and describe some circumstantial evidences. We compute two types of the cylinder amplitudes of open strings and show that these amplitudes approach to the sphere amplitude of closed strings with two winding tachyon insertion in the closed string vacuum limit together with the Hagedorn temperature limit. We also show that the potential energy at the open string vacuum decreases limitlessly as the temperature approaches to the Hagedorn temperature. It is natural to think that the open string vacuum becomes the global minimum near the Hagedorn temperature.}

\begin{document} 
\maketitle
\flushbottom

\section{Introduction}
\label{sec:Intro}

Since the early days of string theory, it was observed that perturbative string gas has an interesting thermodynamic property. The string gas has a characteristic temperature called the Hagedorn temperature \cite{Hagedorn}. The one-loop free energy of strings diverges above this temperature. Over the past decades a considerable number of studies have been made on the thermodynamic behavior of strings near the Hagedorn temperature. In type II string theory, the Hagedorn temperature for closed strings and that for open strings are the same.

In the case of closed string gas, it has been said that the Hagedorn temperature is not really a limiting temperature but rather is associated with a phase transition, in analogy to the deconfining transition in QCD. This is because we can reach the Hagedorn temperature by supplying finite energy in this case \cite{limiting1} \cite{limiting2}. One explanation for the divergence of free energy is that a `winding mode' in the Euclidean time direction in the Matsubara formalism \cite{Matsubara} becomes tachyonic above the Hagedorn temperature. Sathiapalan \cite{Sa}, Kogan \cite{Ko} and Atick and Witten \cite{AW} have proposed the Hagedorn transition of closed strings via condensation of this winding tachyon. Although significant effort has been devoted to search the stable minimum of the potential of this winding tachyon, we have not succeeded in finding it out so far. The potential of closed string tachyon should be calculated on the basis of closed string field theory. However, this theory has not been well-established. That is why it is difficult to find out the potential minimum.

Previously, we have investigated thermodynamic behavior of coincident D-brane--anti-D-brane pairs in type II string theory. Although brane-antibrane pairs are unstable at zero temperature \cite{nonBPSD} (for a review see, e.g., Ref. \cite{Ohmori}), we have pointed out in the previous papers that, in some cases, these unstable branes become stable at sufficiently high temperature \cite{Hotta4} \cite{Hotta5} \cite{Hotta6}.\footnote{For related discussions see Refs. \cite{LowTtach} and \cite{Huang1} \cite{Huang2}.} We have investigated thermodynamic properties of these branes by calculating the finite temperature effective potential of open strings on these branes based on boundary string field theory (BSFT). Let us describe the results, for example, in the case of the brane-antibrane system in a 9-dimensional non-compact flat space. For D9-brane--$\overline{\textrm{D9}}$-brane pairs, a phase transition occurs at slightly below the Hagedorn temperature and the D9-brane--$\overline{\textrm{D9}}$-brane pairs become stable when the temperature increases. On the other hand, for the D$p$-brane--$\overline{\textrm{D} p}$-brane pairs with $p \leq 8$, such a phase transition does not occur. We thus concluded that not a lower dimensional brane-antibrane pairs but D9-brane--$\overline{\textrm{D9}}$-brane pairs are created near the Hagedorn temperature. Let us call this phase transition `{\it thermal brane creation transition}'. This work is applied to cosmology \cite{Hotta7}, and is generalized to the case that a D-brane and an anti-D-brane are separated \cite{Separated}.

There are signs that these two phase transitions, namely, the Hagedorn transition and the thermal brane creation transition are related to each other. The first sign is concerning to the thermodynamic balance of open strings and closed strings on D9-$\overline{\textrm{D9}}$ pairs in type IIB string theory. In the case of an open string gas on the D9-$\overline{\textrm{D9}}$ pairs, we need infinite energy to reach the Hagedorn temperature \cite{limiting1} \cite{limiting2}. This means that the Hagedorn temperature is a `limiting temperature' for open strings on these branes. This is in sharp contrast to the case of closed strings, in which we need finite energy to reach the Hagedorn temperature. Thus, if we consider thermodynamic balance, open strings dominate the total energy of strings \cite{Hotta6}. The open string degrees of freedom are rather important than the closed string ones near the Hagedorn temperature.

The second sign is concerning to world-sheet picture of the Hagedorn transition of closed strings. Atick and Witten argued that, since the insertion of the winding tachyon vertex operator corresponds to the creation of a tiny hole in the world-sheet which wraps around the compactified Euclidean time, an infinite number of tiny holes are created in the process of the winding tachyon condensation \cite{AW}. However, if we identify the boundary of a hole created by winding tachyon vertex operator with a boundary of an open string, we can regard this world-sheet as that of open strings on spacetime-filling branes. In the zero temperature case, Shatashvili conjectured that the closed string world-sheet is obtained from the open string one in the closed string vacuum limit \cite{WorldsheetShatashvili}, and it is examined based on BSFT \cite{WorldsheetAmbJan}, on BCFT \cite{WorldsheetGIR}, on VSFT \cite{WorldsheetGRSZ} and on CSFT \cite{WorldsheetDrukker1} \cite{WorldsheetDrukker2} \cite{WorldsheetTZ} \cite{WorldsheetKKT}.

From these arguments we expect that the stable minimum of the Hagedorn transition of closed strings is related to the open string degrees of freedom. The main purpose of this paper is to present a conjecture that {\it D9-brane--$\overline{\textrm{D9}}$-brane pairs are created by the Hagedorn transition of closed strings,} and describe some circumstantial evidences for this conjecture. We concentrate on D9-brane--$\overline{\textrm{D9}}$-brane pairs in type IIB string theory in this paper, although we can deal with non-BPS D9-branes in type IIA string theory instead.

The outline of the paper is as follows. In \S \ref{sec:HagTra} and \S \ref{sec:DDbar}, we review the Hagedorn transition of closed strings and the thermal brane creation transition, respectively. In \S \ref{sec:Conjecture}, we propose the conjecture mentioned above. Then, we describe the circumstantial evidences for this conjecture. In \S \ref{sec:ClosedStringVacuum}, we compute two types of open string amplitude, namely, the cylinder amplitude and the cylinder amplitude with a single massless boson (massless closed string in NS-NS sector) insertion. We show that each amplitude in the closed string vacuum limit together with the Hagedorn temperature limit approaches to the sphere amplitude for two winding tachyons, and to the sphere amplitude for two winding tachyons and a single massless boson, respectively. It seems reasonable to conclude that the winding tachyon is the closed string vacuum limit of the boundary of an open string. In \S \ref{sec:StableMinimum}, we compare the potential energy at closed string vacuum with that at the open string vacuum. There is good evidence to show that the open string vacuum becomes the potential minimum near the Hagedorn temperature. We conclude in \S \ref{sec:conclusion} with a discussion of future directions. We have also included an appendix, in which we briefly explain the transformation between the one-loop free energy of closed strings in S-representation (E-representation in Ref. \cite{Tan1}) and that in F-representation.

\section{Hagedorn Transition of Closed Strings}
\label{sec:HagTra}

We begin by reviewing the Hagedorn transition of closed strings \cite{Sa} \cite{Ko} \cite{AW}. We can compute the one-loop free energy of strings by using Matsubara method. The one-loop world-sheet of a closed string has torus topology. Then we can derive one-loop free energy in the S-representation (E-representation) or F-representation. The free energy in the F-representation is explicitly invariant under PSL(2,Z) modular transformation of world-sheet torus. However, in order to explain the meaning of winding tachyon, it is convenient to start with the proper time form of one-loop free energy, from which we can directly derive one-loop free energy in S-representation. For ideal gas of many species of point-particles in $(d+1)$-dimensional spacetime, it is given by
\BA
  F (\beta) &=& - \ \frac{v_d}{(2 \pi)^{\frac{d+1}{2}}}
    \int_{0}^{\infty} \frac{ds}{s} \ 
      s^{- \frac{d+1}{2}} \sum_{M_b} \sum_{w=1}^{\infty}
        \exp \left( - \ \frac{M_b^2}{2} \ s
          - \frac{w^2 \beta^2}{2s} \right)
            \nonumber \\
  && + \frac{v_d}{(2 \pi)^{\frac{d+1}{2}}}
    \int_{0}^{\infty} \frac{ds}{s} \ 
      s^{- \frac{d+1}{2}} \sum_{M_f} \sum_{w=1}^{\infty}
        (-1)^w \exp \left( - \ \frac{M_f^2}{2} \ s
          - \frac{w^2 \beta^2}{2s} \right),
\label{eq:propertime}
\EA
where $M_b$ represents the mass of boson, and $M_f$ that of fermion, and we take the sum over mass eigenvalues. $w$ is the number of times which the world line wind around the compactified Euclidean time. We can compute the one-loop free energy of superstrings as the collection of point-particles with superstring mass spectrum.

\begin{figure}[tbp]
\centering
\includegraphics[width=.3\textwidth]{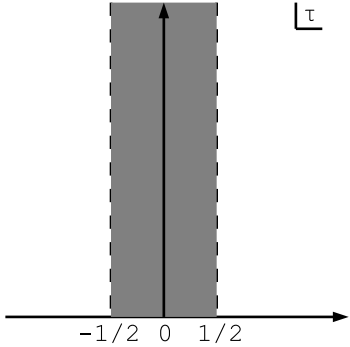}
\caption{\label{fig:Srep} Shaded region is the domain of integration ${\cal S}$ in the S-representation.}
\end{figure}

\begin{figure}[tbp]
\centering
\includegraphics[width=.3\textwidth]{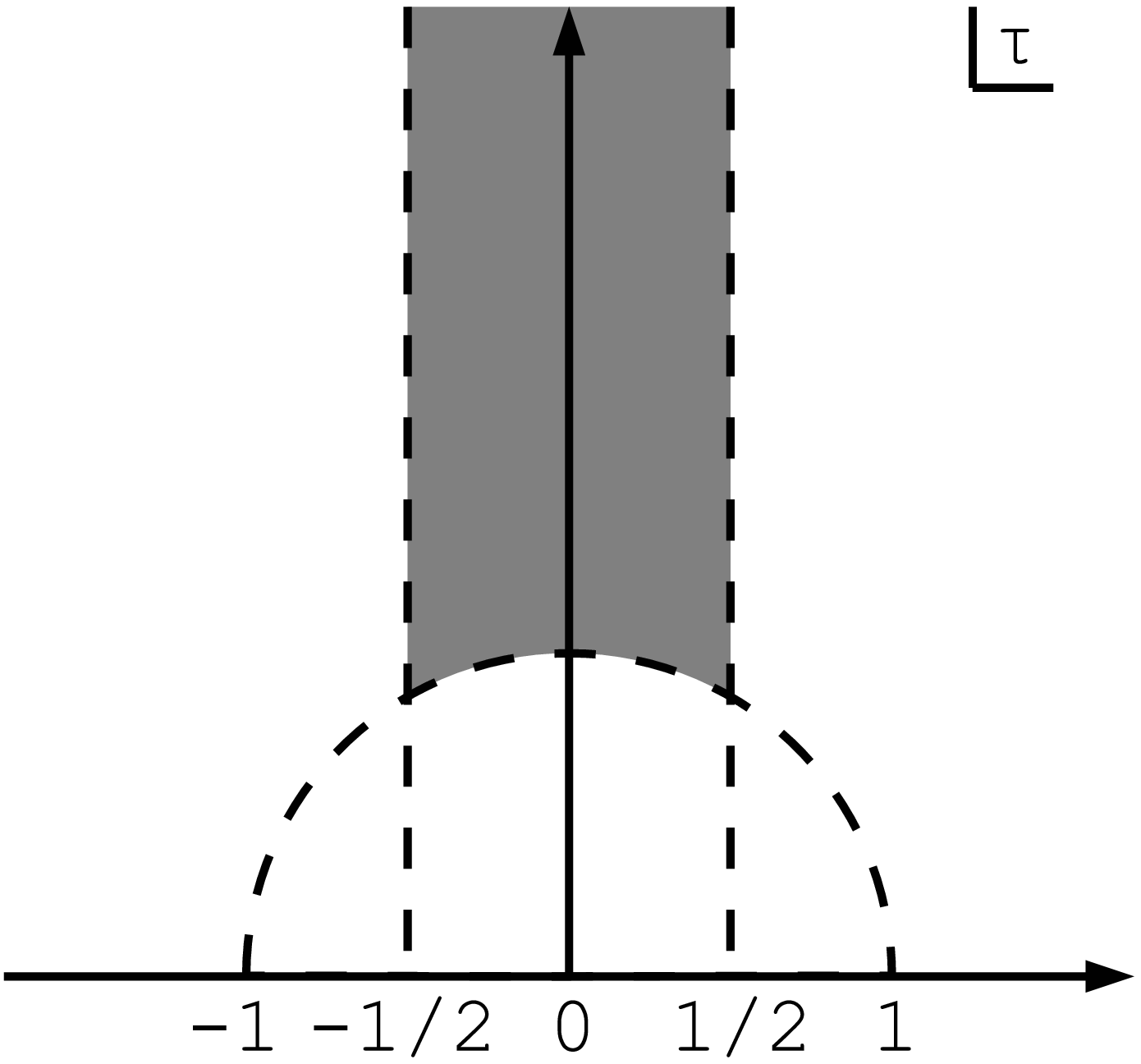}
\caption{\label{fig:Frep} Shaded region is the domain of integration ${\cal F}$ in the F-representation.}
\end{figure}

We consider the type IIB closed string gas in 10-dimensional spacetime. For the closed string gas, it is convenient to rewrite (\ref{eq:propertime}) as
\BA
  F (\beta) &=& - \ \frac{2^5 v_9}{\beta_H^{10}}
    \int_{0}^{\infty} \frac{d \tau_2}{\tau_2^6}
      \sum_{M_b} \sum_{w=1}^{\infty}
        \exp \left( - \pi \alpha ' M_b^2 \tau_2
          - \frac{2 \pi w^2 \beta^2}{\beta_H^2 \tau_2} \right)
            \nonumber \\
  && + \frac{2^5 v_9}{\beta_H^{10}}
    \int_{0}^{\infty} \frac{d \tau_2}{\tau_2^6}
      \sum_{M_f} \sum_{w=1}^{\infty}
        (-1)^w \exp \left( - \pi \alpha ' M_f^2 \tau_2
          - \frac{2 \pi w^2 \beta^2}
            {\beta_H^2 \tau_2} \right),
\label{eq:closedpropertime}
\EA
by using the variable $\tau_2$ defined as
\BA
  s = 2 \pi \alpha ' \tau_2.
\label{eq:closedstau2}
\EA
$\alpha '$ is the slope parameter and $\beta_H$ the inverse of the Hagedorn temperature
\BE
  \beta_H = 2 \pi \sqrt{2 \alpha '}.
\label{eq:BH}
\EE
In this case, $w$ is the number of times which the world-sheet wind around the compactified Euclidean time. The mass square of closed superstrings in the light-cone gauge is given by
\BA
  M^2 = \frac{2}{\alpha '} \left( N - \nu + \tilde{N} - \tilde{\nu} \right),
\EA
where $N$ and $\tilde{N}$ denote the oscillation numbers of left-moving mode and right-moving mode, respectively. These oscillation numbers depend on the periodicity condition of fermions on the world-sheet. The oscillation number of the left-moving mode is given by
\BA
  N = N_B + N_{NS},
\label{eq:oscNS}
\EA
for Neveu-Schwarz (NS) condition, and
\BA
  N = N_B + N_R,
\label{eq:oscR}
\EA
for Ramond (R) condition, and the same for the right-moving mode. $N_B$, $N_{NS}$ and $N_R$ are ordinary written as operators. However, we need only their eigenvalues here, and they are given by
\BA
  N_B &=& \sum_{I=2}^9 \sum_{l=1}^{\infty} l n_l^I, \\
  N_{NS}
    &=& \sum_{I=2}^9 \sum_{r = \frac{1}{2}}^{\infty} r n_r^I, \\
  N_R &=& \sum_{I=2}^9 \sum_{m=1}^{\infty} m n_m^I,
\EA
where $n_l^I = 0, 1, 2, \cdots$ and $n_r^I, n_m^I = 0, 1$. $\nu$ and $\tilde{\nu}$ are $1/2$ for NS mode, and $0$ for R mode. The eigenvalues of the GSO projection operators for the left-moving mode are given by
\BA
  P_{NS} &=& \frac{1}{2} \left\{ 1
    - (-1)^{\sum_{I=2}^9 \sum_{r = \frac{1}{2}}^{\infty} n_r^I}
      \right\}, \\
  P_R &=& \frac{1}{2} \left\{ 1
    + (-1)^{\sum_{I=2}^9 \sum_{m=1}^{\infty} n_m^I
      + \sum_{a=1}^4 n_a} \right\},
\EA
and the same for the right-moving mode. $n_a$ is related to the zero modes of R fermion and $n_a = 0, 1$. In order to impose level-matching condition
\BA
  \left( N - \nu \right) - \left( \tilde{N} - \tilde{\nu} \right) = 0,
\label{eq:constraint}
\EA
we insert the integral representation of Kronecker-delta,
\BA
  \delta_{N - \nu , \tilde{N} - \tilde{\nu}}
    = \int_{- \frac{1}{2}}^{\frac{1}{2}} d \tau_1 \ 
      \exp \left[ 2 \pi i \tau_1
        \left( N - \nu - \tilde{N} + \tilde{\nu} \right) \right],
\EA
to the summation over the mass eigenvalues in equation (\ref{eq:closedpropertime}). Taking these conditions into consideration, we obtain the one-loop free energy in the S-representation as\footnote{We can add $w=0$ terms in the braces, since these terms cancel out due to the identity (\ref{eq:aequatioidentica}).}
\BA
  F_c (\beta)
    &=& - \ \frac{8 (2 \pi)^8 v_9}{\beta_H^{10}}
      \int_{\cal S} \frac{d^2 \tau}{\tau_2^6}
        \frac{1}{\left| {\vartheta_1}' (0 | \tau) \right|^8}
          \nonumber \\
  && \hspace{3mm}
    \times \left[ \left\{ \left( \vartheta_3^4 - \vartheta_4^4 \right)
      \left( {\bar{\vartheta}}_3^4 - {\bar{\vartheta}}_4^4 \right)
        + \vartheta_2^4 {\bar{\vartheta}}_2^4 \right\} (0 | \tau) \ 
          \sum_{w=1}^{\infty}
            \exp \left(- \ \frac{2 \pi w^2 \beta^2}
              {\beta_H^2 \tau_2} \right)
                \right. \nonumber \\
  && \hspace{3mm}
    \left. - \left\{ \left( \vartheta_3^4 - \vartheta_4^4 \right)
      {\bar{\vartheta}}_2^4 + \vartheta_2^4
        \left( {\bar{\vartheta}}_3^4
          - {\bar{\vartheta}}_4^4 \right) \right\} (0 | \tau) \ 
            \sum_{w=1}^{\infty} (-1)^w
              \exp \left(- \ \frac{2 \pi w^2 \beta^2}
                {\beta_H^2 \tau_2} \right)
                  \right], \nonumber \\
\label{eq:closedSrep}
\EA
where $\tau = \tau_1 + i \tau_2$, and the domain of integration ${\cal S}$ is sketched in Figure \ref{fig:Srep}. Comparing this with the results of path integral calculation, we can identify $\tau$ as moduli parameter of torus world-sheet. However, this free energy is not explicitly invariant under PSL(2,Z) modular transformation, which is generated by the shift $T$ and the inversion $S$ defined as
\BA
  T : \ \ \ \tau &\rightarrow& \tau + 1,
\label{eq:modular1} \\
  S : \ \ \ \tau &\rightarrow& - \ \frac{1}{\tau}.
\label{eq:modular2}
\EA
We can transform the free energy into modular invariant form by using unfolding technique \cite{Tan1} \cite{McRoth}, as we shortly see in the Appendix A. Then we obtain the one-loop free energy in the F-representation as \cite{AW}
\BA
  F_c (\beta) &=& - \ \frac{8 (2 \pi)^8 v_9}
    {\beta_H^{10}}
      \int_{\cal F} \frac{d^2 \tau}{\tau_2^6}
        \frac{1}{\left| {\vartheta_1}' (0 | \tau) \right|^8} \ 
          {\sum_{m,n = - \infty}^{\infty}}' e^{- S_{\beta} (m,n)}
            \nonumber \\
  && \hspace{10mm}
    \times \left\{ \left( \vartheta_2^4 {\bar{\vartheta}}_2^4
      + \vartheta_3^4 {\bar{\vartheta}}_3^4
        + \vartheta_4^4 {\bar{\vartheta}}_4^4 \right) (0 | \tau)
          + (-1)^{m+n} \left( \vartheta_2^4 {\bar{\vartheta}}_4^4
            + \vartheta_4^4 {\bar{\vartheta}}_2^4 \right) (0 | \tau)
              \right. \nonumber \\
  && \hspace{15mm} \left.
    - (-1)^n \left( \vartheta_2^4 {\bar{\vartheta}}_3^4
      + \vartheta_3^4 {\bar{\vartheta}}_2^4 \right) (0 | \tau)
        - (-1)^m \left( \vartheta_3^4 {\bar{\vartheta}}_4^4
          + \vartheta_4^4 {\bar{\vartheta}}_3^4 \right) (0 | \tau)
            \right\},
\label{eq:closedFrep}
\EA
where we have defined
\BA
  S_{\beta} (m,n) = \frac{2 \pi \beta^2}{\beta_H^2 \tau_2}
    \left( n^2 + m^2 |\tau| + 2 \tau_1 mn \right).
\EA
The domain of integration ${\cal F}$ is one of the fundamental regions of modular transformation, and is sketched in Figure \ref{fig:Frep}. The prime over the sum indicates the exclusion of $m=n=0$. We can add $m=n=0$ term and get rid of the prime over the sum, since the term vanishes due to well-known {\it aequatio identica satis abstrusa}
\BE
  \vartheta_2^4 (0 | \tau) - \vartheta_3^4 (0 | \tau)
    + \vartheta_4^4 (0 | \tau) = 0.
\label{eq:aequatioidentica}
\EE
The free energy in this expression is explicitly modular invariant \cite{AW}. In order to see this, it is sufficient to prove the invariance under the transformations (\ref{eq:modular1}) and (\ref{eq:modular2}). For the shift $T$ (\ref{eq:modular1}), we can verify it by using the modular transformation of the Jacobi theta functions
\BA
  && {\vartheta_1}' ( 0 | \tau )
    = e^{\frac{\pi i}{4}} {\vartheta_1}' ( 0 | \tau + 1 ), \\
  && \vartheta_2 ( 0 | \tau )
    = e^{\frac{\pi i}{4}} \vartheta_2 ( 0 | \tau + 1 ), \\
  && \vartheta_3 ( 0 | \tau )
    = \vartheta_4 ( 0 | \tau + 1 ), \\
  && \vartheta_4 ( 0 | \tau )
    = \vartheta_3 ( 0 | \tau + 1 ),
\EA
and replacing $m' = m$ and $n' = m+n$, and for the inversion $S$ (\ref{eq:modular2}), by using
\BA
  && {\vartheta_1}' ( 0 | \tau )
    = e^{\frac{3 \pi i}{4}} \tau^{- \frac{3}{2}}
      {\vartheta_1}' \left( 0 \left| - \frac{1}{\tau} \right. \right),
\label{eq:thetaS1} \\
  && \vartheta_2 ( 0 | \tau )
    = e^{\frac{\pi i}{4}} \tau^{- \frac{1}{2}}
      \vartheta_4 \left( 0 \left| - \frac{1}{\tau} \right. \right), \\
  && \vartheta_3 ( 0 | \tau )
    = e^{\frac{\pi i}{4}} \tau^{- \frac{1}{2}}
      \vartheta_3 \left( 0 \left| - \frac{1}{\tau} \right. \right), \\
  && \vartheta_4 ( 0 | \tau )
    = e^{\frac{\pi i}{4}} \tau^{- \frac{1}{2}}
      \vartheta_2 \left( 0 \left| - \frac{1}{\tau} \right. \right),
\label{eq:thetaS4}
\EA
and replacing $m' = -n$ and $n' = m$. Let us denote the greatest common divisor of integers $x$ and $y$ as $[x,y]$. Then $w$ in (\ref{eq:closedSrep}) and $m$, $n$ in (\ref{eq:closedFrep}) are related as $[m,n] = w$ as is explained in Ref. \cite{Tan1} and \cite{McRoth}, as we see in Appendix A.

In order to see the winding tachyon, we further transform this one-loop free energy by using the Poisson resummation formula
\BA
  \sum_{n = - \infty}^{\infty} f(n)
    = \sum_{j = - \infty}^{\infty}
      \int_{- \infty}^{\infty} du \ \exp (2 \pi i j u) f(u).
\label{eq:PoissonResum}
\EA
For example, for the first term in the braces in (\ref{eq:closedFrep}), the summation over $m$ and $n$ is transformed as
\BA
  \sum_{m,n = - \infty}^{\infty} e^{- S_{\beta} (m,n)}
    &=& \sum_{m,n = - \infty}^{\infty}
      \exp \left[ - \left\{ n^2 + 2 m n \tau_1
        + m^2 (\tau_1^2 + \tau_2^2) \right\}
          \frac{2 \pi \beta^2}{\beta_H^2 \tau_2} \right]
            \nonumber \\
  &=& \sum_{m,j = - \infty}^{\infty} \int_{- \infty}^{\infty} du \ 
    \exp (2 \pi i j u) \exp \left[ - \left\{ u^2 + 2 m u \tau_1
      + m^2 (\tau_1^2 + \tau_2^2) \right\}
        \frac{2 \pi \beta^2}{\beta_H^2 \tau_2} \right]
          \nonumber \\
  &=& \sum_{m,j = - \infty}^{\infty} \int_{- \infty}^{\infty} du \ 
    \exp \left[ - \left\{ \left( u + m \tau_1
      - \ \frac{i j \beta_H^2 \tau_2}{\beta^2} \right)^2
        \right. \right. \nonumber \\
  && \hspace{40mm} \left. \left.
    + \frac{j^2 \beta_H^4 \tau_2^2}{4 \beta^4}
      + \frac{i m j \beta_H^2 \tau_1 \tau_2}{\beta^2}
        + m^2 \tau_2^2 \right\}
          \frac{2 \pi \beta^2}{\beta_H^2 \tau_2} \right]
            \nonumber \\
  &=& \frac{\beta_H}{\beta} \sqrt{\frac{\tau_2}{2}}
    \sum_{m,j = - \infty}^{\infty}
      \exp \left[ - 2 \pi
        \left( \frac{j^2 \beta_H^2 \tau_2}{4 \beta^2}
          + i m j \tau_1 + \frac{m^2 \beta^2 \tau_2}
            {\beta_H^2} \right) \right].
\EA
We can also transform other terms similarly, if we notice that, for example, $(-1)^m = e^{\pi i m}$. Then, we obtain the one-loop free energy in dual-representation as \cite{AW}
\BA
  F_c (\beta)
    &=& - \ \frac{4 \sqrt{2} (2 \pi)^8 v_9}
      {\beta_H^9 \beta}
        \int_{\cal F} \frac{d^2 \tau}{\tau_2^{\frac{11}{2}}}
          \frac{1}{\left| {\vartheta_1}' (0 | \tau) \right|^8}
            \nonumber \\
  && \hspace{30mm}
    \times \left[ \left( \vartheta_2^4 {\bar{\vartheta}}_2^4
      + \vartheta_3^4 {\bar{\vartheta}}_3^4
        + \vartheta_4^4 {\bar{\vartheta}}_4^4 \right) (0 | \tau)
          \sum_{m,j} D_1 (m , j , \beta ; \tau)
            \right. \nonumber \\
  && \hspace{35mm} + \left( \vartheta_2^4 {\bar{\vartheta}}_4^4
    + \vartheta_4^4 {\bar{\vartheta}}_2^4 \right) (0 | \tau)
      \sum_{m,j}
        D_2 \left( m , j + \frac{1}{2} , \beta ; \tau \right)
          \nonumber \\
  && \hspace{35mm}
    - \left( \vartheta_2^4 {\bar{\vartheta}}_3^4
      + \vartheta_3^4 {\bar{\vartheta}}_2^4 \right) (0 | \tau)
        \sum_{m,j}
          D_1 \left( m , j + \frac{1}{2} , \beta ; \tau \right)
            \nonumber \\
  && \hspace{35mm} \left.
    - \left( \vartheta_3^4 {\bar{\vartheta}}_4^4
      + \vartheta_4^4 {\bar{\vartheta}}_3^4 \right) (0 | \tau)
        \sum_{m,j} D_2 (m , j , \beta ; \tau)
          \right],
\EA
where we have defined
\BA
  D_1 (m , j , \beta ; \tau)
    &=& \exp \left[ - S_{\beta}^D (m , j ; \tau) \right], \\
  D_2 (m , j , \beta ; \tau)
    &=& (-1)^m \exp \left[ - S_{\beta}^D (m , j ; \tau) \right], \\
  S_{\beta}^D (m , j ; \tau)
    &=& 2 \pi \left( m^2 \ \frac{\beta^2}
      {\beta_H^2} \ \tau_2 + i mj \tau_1
        + j^2 \ \frac{\beta_H^2}{4 \beta^2} \ 
          \tau_2 \right).
\EA
The form of $S_{\beta}^D (m , j ; \tau)$ is reminiscent of T-duality \cite{Tdual1} \cite{Tdual2}. However, this one-loop free energy is not invariant under the exchange of $\sqrt{2} \beta / \beta_H$ and $\beta_H / (\sqrt{2} \beta)$.\footnote{In heterotic string theory, one-loop free energy is invariant under this type of transformation \cite{Tan1}.} If we focus on $m = \pm 1,$ $j = 0$ part in $(\vartheta_3^4 {\bar{\vartheta}}_3^4 + \vartheta_4^4 {\bar{\vartheta}}_4^4)$ term and $(\vartheta_3^4 {\bar{\vartheta}}_4^4 + \vartheta_4^4 {\bar{\vartheta}}_3^4)$ term, we obtain a factor $\exp [- 2 \pi (\beta^2 - \beta_H^2) \tau_2 / \beta_H^2]$.\footnote{If we focus on $m = 0,$ $j = \pm 1$ part instead, these terms cancel each other.} It diverges as $\tau_2 \rightarrow \infty$ if $\beta < \beta_H$. This is the origin of the divergence of the one-loop free energy above the Hagedorn temperature. Put it another way, comparing with the proper time form of one-loop free energy (\ref{eq:closedpropertime}), we can express the `mass' of these modes as
\BA
  M^2 = \frac{2}{\alpha '} \ \frac{\beta^2 - \beta_H^2}{\beta_H^2},
\label{eq:windtacmass}
\EA
and they become tachyonic for $\beta < \beta_H$. These modes are called winding tachyon. It is noteworthy that this winding tachyon corresponds to the closed string winding once around the compactified Euclidean time. Since $[\pm 1, l] = 1$ for an arbitrary integer $l$, the $m = \pm 1$ terms come from $w=1$ terms in (\ref{eq:closedSrep}), and corresponding world-sheet winds around the compactified Euclidean time once. Sathiapalan \cite{Sa}, Kogan \cite{Ko} and Atick and Witten \cite{AW} have proposed that the phase transition occurs via the condensation of these tachyonic modes. Atick and Witten further argued about the world-sheet picture of strings \cite{AW}. The insertion of the winding tachyon vertex operator means the creation of a tiny hole in the world-sheet which wraps around the compactified Euclidean time. Thus, the addition of the winding tachyon vertex operators to the world-sheet action induces the creation of a sea of such holes. At low temperature, sphere world-sheet does not contribute to the free energy, since it cannot wrap the compactified Euclidean time. However, if we consider the condensation of winding tachyon above the Hagedorn temperature, the sphere world-sheet is no longer simply connected, and it contributes to the free energy above the Hagedorn temperature. From the calculation of the coupling of a dilaton and two winding tachyons they inferred that the Hagedorn transition is first order phase transition, and that it takes place at slightly below the Hagedorn temperature.

It should be noted that these modes can be interpreted as winding tachyon only in the Matsubara formalism, namely, if we perform the Wick rotation of the time direction and compactified it with period $\beta$. As we can see from above calculation, the winding tachyon is derived by regarding temperature dependent part as mass part in the free energy. We cannot identify which modes condensate to what extent in Lorentzian time when this winding tachyon condensates in the Euclidean time.

\section{Thermal Brane Creation Transition}
\label{sec:DDbar}

In this section we review our previous work about coincident D-brane--anti-D-brane pairs at finite temperature \cite{Hotta4} \cite{Hotta5} \cite{Hotta6}. We have investigated the thermodynamic properties of the D-brane--anti-D-brane pairs by calculating the finite temperature effective action of open strings on these branes. We can evaluate the tachyon potential for a D-brane--anti-D-brane pair by using BSFT \cite{BSFT1} \cite{BSFT2} \cite{BSFT3} \cite{BSFT4}. This theory is based on the Batalin-Vilkoviski formalism \cite{BV}. The effective action for tachyon field is provided by the master equation of this formalism. In the case of superstrings, the solution of the classical master equation is given by \cite{tachyon2} {tachyon3} \cite{tachyon4} \cite{TakaTeraUe}
\BE
  I_{eff} = Z,
\label{eq:SoZo}
\EE
where $I_{eff}$ is the spacetime effective action and $Z$ is the disk partition function of the two-dimensional world-sheet theory. If we calculate this solution in a constant tachyon background in the case of a single D$p$-$\overline{\textrm{D$p$}}$ pair in type II string theory, we obtain the tachyon potential \cite{tachyon2} \cite{TakaTeraUe} \cite{tachyon1}
\BE
  V(T) = 2 \tau_p v_p \exp (-8 |T|^2),
\EE
where $T$ is a complex scalar tachyon field and $v_p$ is the $p$-dimensional volume of the system that we are considering. $\tau_p$ is the tension of a D$p$-brane, which is defined by
\BE
  \tau_p = \frac{1}
    {(2 \pi)^p {\alpha '}^{\scriptscriptstyle \frac{p+1}{2}} g_s},
\label{eq:tension}
\EE
where $g_s$ is the coupling constant of strings. The potential has the maximum at $T=0$ and has the minimum at $|T| = \infty$ in this case. We can adjust the overall coefficient such that this potential satisfies Sen's conjecture \cite{Senconjecture1} \cite{Senconjecture2} \cite{Senconjecture3} \cite{Senconjecture4}, that is, the potential height equals to the tension of the D$p$-$\overline{\textrm{D$p$}}$ pair.

In order to calculate the free energy by using the Matsubara method in the ideal gas approximation, we must calculate the one-loop amplitude. We have assumed that the relation (\ref{eq:SoZo}) also holds if we include the higher loop correction of the world-sheet theory, namely, if $Z$ is the partition function to all orders of the two-dimensional world-sheet theory. If we consider the one-loop amplitude based on BSFT, we are confronted with the problem of the choice of the Weyl factors \cite{1loopAO} \cite{1loopann1} \cite{1loopann2} \cite{1loopsym1} \cite{1loopsym2} \cite{1loop1} \cite{1loop2} \cite{1loop3} \cite{1loop4}. Andreev and Oft have proposed the following form of boundary action in the Minkowski spacetime in type II string theory \cite{1loopAO}:
\BE
  I_{\partial \Sigma}
    = \int_{0}^{2 \pi \tau} d \sigma^2 \int_{0}^{\pi} d \sigma^1
      \left[ |T|^2 \delta (\sigma^1)
        + |T|^2 \delta (\pi - \sigma^1) \right],
\label{eq:BoundaryAction}
\EE
where we have used the Euclidean convention \cite{Polchinski}. This action is natural in the sense that both sides of the cylinder world-sheet are treated on an equal footing in this case. They proposed this action on the basis of the principle that low energy part of one-loop free energy should coincide with that of the tachyon field model \cite{tachyon2} \cite{tachyon3} \cite{TakaTeraUe} \cite{tachyon1}.

We have investigated D$p$-$\overline{\textrm{D$p$}}$ pairs in a 9-dimensional non-compact flat space \cite{Hotta4}. If we compute the one-loop amplitude in the space where the Euclidean time direction is compactified with period $\beta$, we obtain the free energy. We will explain the precise derivation in the case of a D$p$-$\overline{\textrm{D$p$}}$ pair in \S \ref{sec:ClosedStringVacuum}. The result is
\BA
  F_o (T, \beta)
    &=& - \ \frac{2 (2 \pi)^4 v_p}{\beta_H^{p+1}}
      \int_{0}^{\infty} \frac{d \tau}{\tau} \ 
        \tau^{- \frac{p+1}{2}} e^{-4 \pi |T|^2 \tau}
          \nonumber \\
  && \hspace{20mm}
    \times \left[ \left\{ \frac{\vartheta_3 (0 | i \tau)}
      {{\vartheta_1}' (0 | i \tau)} \right\}^4
        \left\{ \sum_{w=1}^{\infty}
          \exp \left( - \frac{\pi w^2 \beta^2}{\beta_H^2 \tau}
            \right) \right\}
              \right. \nonumber \\
  && \hspace{30mm} \left.
    - \left\{ \frac{\vartheta_2 (0 | i \tau)}
      {{\vartheta_1}' (0 | i \tau)} \right\}^4
        \left\{ \sum_{w=1}^{\infty} (-1)^w
          \exp \left( - \frac{\pi w^2 \beta^2}{\beta_H^2 \tau}
            \right) \right\} \right].
\label{eq:Dp_antiDp_Free}
\EA
This free energy can also be obtained from the proper time form of the free energy \cite{Pol1loop} \cite{Alvarez} \cite{AlvOso}. For open string gas on D$p$-$\overline{\textrm{D$p$}}$ pairs, it is convenient to express (\ref{eq:propertime}) as
\BA
  F (\beta) &=& - \ \frac{v_p}{\beta_H^{p+1}}
    \int_{0}^{\infty} \frac{d \tau}{\tau^{\frac{p+3}{2}}}
      \sum_{M_b} \sum_{w=1}^{\infty}
        \exp \left( - 2 \pi \alpha ' M_b^2 \tau
          - \frac{\pi w^2 \beta^2}{\beta_H^2 \tau} \right)
            \nonumber \\
  && + \frac{v_p}{\beta_H^{p+1}}
    \int_{0}^{\infty} \frac{d \tau}{\tau^{\frac{p+3}{2}}}
      \sum_{M_f} \sum_{w=1}^{\infty}
        (-1)^w \exp \left( - 2 \pi \alpha ' M_f^2 \tau
          - \frac{\pi w^2 \beta^2}
            {\beta_H^2 \tau} \right),
\EA
by using the variable $\tau$ defined as
\BA
  s = 4 \pi \alpha ' \tau.
\label{eq:openstau}
\EA
The difference of definition of integration variables (\ref{eq:closedstau2}) and (\ref{eq:openstau}) cause the difference in the power of exponential. $w$ is the number of times which the world-sheet wind around the compactified Euclidean time, like in the case of closed strings. We can derive the free energy (\ref{eq:Dp_antiDp_Free}) by substituting the mass square in the light-cone gauge \cite{Hotta4}
\BA
  M^2 &=& \frac{1}{\alpha '} \left( N + 2 |T|^2 - \nu \right),
\EA
where the oscillation number is given by (\ref{eq:oscNS}) for NS condition and (\ref{eq:oscR}) for R condition. $\nu$ is $1/2$ for NS mode, and $0$ for R mode like in the closed string case. If we consider multiple D-brane--anti-D-brane pairs, for example, $N$ D$p$-$\overline{\textrm{D$p$}}$ pairs, we need to replace the tachyon field $T$ to the $(N, \overline{N})$ representation of the $U(N) \times U(N)$ gauge group, which we denote by ${\bf T}$ \cite{IIBornotIIB}. In this case, the tachyon potential is given by
\BE
  V ({\bf T}) = 2 \tau_p v_p \ {\rm Tr}
    \exp \left( -8 {{\bf T}}^{\dagger} {\bf T} \right),
\label{eq:TMNpotential}
\EE
and it has the maximum at ${\bf T} = {\bf 0}$ (${\bf 0}$ is the $N \times N$ zero matrix). We have explicitly calculated the finite temperature effective potential in the case that ${\bf T} = \textrm{diag} (T \cdots T)$. We have computed it by using the microcanonical ensemble method, since we cannot trust the canonical ensemble method near the Hagedorn temperature in general \cite{Efura1} \cite{Efura2}. One may notice that, in the D$9$-$\overline{\textrm{D$9$}}$ pair case which we are most interested in, two methods lead to the same results \cite{Hotta4}.

We summarize here the results of our previous works about the finite temperature systems of D-brane--anti-D-brane pairs in a constant tachyon background \cite{Hotta4} \cite{Hotta5} \cite{Hotta6}. We have investigated D$p$-$\overline{\textrm{D$p$}}$ pairs in a 9-dimensional non-compact flat space. The result of the D9-$\overline{\textrm{D9}}$ pair case is in sharp contrast to that of the D$p$-$\overline{\textrm{D$p$}}$ pair case with $p \leq 8$. In the case of $N$ D9-$\overline{\textrm{D9}}$ pairs, the $|T|^2$ term of the finite temperature effective potential near the Hagedorn temperature is approximated as
\BE
  \left[ -16 N \tau_9 v_9
   + \frac{8 \pi N^2 v_9}{\beta_H^{10}}
    \ln \left( \frac{\pi \beta_H^{10} E}
      {2 N^2 v_9} \right) \right] |T|^2,
\label{eq:p9T2E}
\EE
where $E$ is the energy of open strings. We must impose the condition that the 't Hooft coupling is very small, namely,
\BE
  g_s N \ll 1,
\label{eq:gN}
\EE
for the ideal gas approximation. Because the first term in the coefficient of $|T|^2$ is a constant as long as $v_9$ and $\tau_9$ are fixed, and the second term is an increasing function of $E$, the sign of the $|T|^2$ term changes from negative to positive as $E$ increases. The coefficient vanishes at the critical temperature ${\cal T}_c$ which is given by
\BE
  {\cal T}_c
    \simeq \beta_H^{-1}
      \left[ 1 + \exp \left(
        - \frac{\beta_H^{10} \tau_9}{\pi N}
          \right) \right]^{-1}.
\label{eq:Tc}
\EE
This implies that ${\bf T} = {\bf 0}$ is the potential minimum above the critical temperature. From this we can see that a phase transition occurs at the critical temperature ${\cal T}_c$, which is slightly below the Hagedorn temperature, and the D9-$\overline{\textrm{D9}}$ pairs are stable above this critical temperature. The total energy at the critical temperature is a decreasing function of $N$ as long as the 't Hooft coupling is very small in these cases. This implies that a large number $N$ of D9-$\overline{\textrm{D9}}$ pairs are created simultaneously. In order to understand the precise thermodynamic behavior, we need to perform a non-perturbative calculation. For the D$p$-$\overline{\textrm{D$p$}}$ pairs with $p \leq 8$, on the other hand, the coefficient remains negative near the Hagedorn temperature, so that such a phase transition does not occur. We thus concluded that, in type IIB string theory, not lower dimensional D$p$-$\overline{\textrm{D$p$}}$ pairs but D9-$\overline{\textrm{D9}}$ pairs are created near the Hagedorn temperature. This is the thermal brane creation transition which we mentioned in \S \ref{sec:Intro}.

We have also investigated D$p$-$\overline{\textrm{D$p$}}$ pairs in a toroidal space. We have supposed that $D$-dimensional space is toroidally compactified and that the rest of the $(9-D)$-dimensional space is left uncompactified ($M_{1,9-D} \times T_{D}$). We have also assumed that the D$p$-$\overline{\textrm{D$p$}}$ pairs extend in the $d$-dimensional space in the non-compact direction, and in the $(p-d)$-dimensional space in the toroidal direction. If $D+d=9$, that is, the D$p$-$\overline{\textrm{D$p$}}$ pairs are extended in all the non-compact directions, the thermal brane creation transition occurs near the Hagedorn temperature and these branes become stable. On the other hand, if $D+d \leq 8$, that is, the D$p$-$\overline{\textrm{D$p$}}$ pairs are not extended in all the non-compact directions, the thermal brane creation transition does not occur. It is noteworthy that spacetime-filling branes, D9-$\overline{\textrm{D9}}$ pairs in the type IIB case, are created at sufficiently high energy not only in a non-compact space but also in a toroidal space, since these branes always satisfy $D+d=9$.

\section{Conjecture}
\label{sec:Conjecture}

In this section we contemplate the relationship between two phase transitions, that is, the Hagedorn transition of closed strings and the thermal brane creation transition. Let us consider the situation that there are closed strings together with open strings in type IIB string theory. Previously we studied the thermodynamic balance between closed strings and open strings in the presence of D9-$\overline{\textrm{D9}}$ pairs in the ideal gas approximation near the Hagedorn temperature \cite{Hotta6}. In this case, energy flows from closed strings to open strings, and open strings dominate the total energy. This is because we can reach the Hagedorn temperature for closed strings by supplying finite energy, while we need infinite energy to reach the Hagedorn temperature for open strings on these branes. This implies that, as the temperature increases, the creation of D9-$\overline{\textrm{D9}}$ pairs begin before closed strings are highly excited.

Let us return to the world-sheet picture of winding tachyon we have described in \S \ref{sec:HagTra}. Atick and Witten argued about the meaning of the condensation of the winding tachyon \cite{AW}. The insertion of the winding tachyon vertex operator corresponds to the creation of a tiny hole in the world-sheet which wraps around the compactified Euclidean time, and the condensation of winding tachyon induces an infinite number of tiny holes in the world-sheet. However, what is the hole of closed string world-sheet? Let us try to think about it from a different point of view. Taking a time slice of a sphere world-sheet with some winding tachyon insertion, we obtain open strings because the boundaries of holes wind around the Euclidean time direction. Therefore, this world-sheet represents open strings propagating in the Euclidean time direction. Then we can identify the boundary of a hole created by winding tachyon vertex operator with a boundary of an open string on a D9-$\overline{\textrm{D9}}$ pair, and the insertion of winding tachyon vertex operator means the insertion of the boundary of open strings in the tiny hole limit, which wraps the compactified Euclidean time once. Then the sphere world-sheet with multiple winding tachyon vertex operator insertion is naturally reinterpreted as multi-loop world-sheet of open strings. If we enlarge the size of this hole, we can describe open strings with arbitrary boundary. Therefore, we present the following conjecture in type IIB string theory :

\begin{center}

\framebox{\it D9-brane--$\overline{\textrm{D9}}$-brane pairs are created by the Hagedorn transition of closed strings.}

\end{center}

That is to say, above two phase transitions are two aspects of one phase transition. In the sense that ${\bf T} = {\bf 0}$ is the perturbative vacuum of open strings, this is a phase transition from the closed string vacuum to the open string vacuum. In other words, the open string vacuum becomes the stable minimum of the Hagedorn transition near the Hagedorn temperature. In the following two sections we describe some circumstantial evidences for this conjecture.

\section{Correspondence in the Closed String Vacuum Limit}
\label{sec:ClosedStringVacuum}

In this section, we consider two types of open string amplitude close to the closed string vacuum near the Hagedorn temperature. We show that the cylinder amplitude for large $|T|$ has the form of the propagator of winding tachyon. We also show that the cylinder amplitude with a single massless boson insertion approaches to the sphere amplitude with two winding tachyons and a single massless boson insertion by taking the closed string vacuum limit together with the Hagedorn temperature limit appropriately. This enables us that we can identify a vertex operator of winding tachyon as the closed string vacuum limit of the boundary of an open string world-sheet, which wraps once around the compactified Euclidean time.

\subsection{Cylinder World-sheet}
\label{sec:Propagator}

Let us consider the cylinder amplitude of open strings on a D9-$\overline{\textrm{D9}}$ pair. In the ideal gas approximation, free energy of open strings can be obtained by calculating cylinder amplitude in the space where the Euclidean time direction is compactified with period $\beta$. In the case of open string gas on a BPS D-brane, it was shown that the one-loop free energy of open strings can be transformed to the propagator of a closed string \cite{VM}. We here perform similar calculation in the case of open string gas on a D9-$\overline{\textrm{D9}}$ pair. As we have mentioned in \S \ref{sec:DDbar}, we can compute the one-loop free energy of open strings on a D$p$-$\overline{\textrm{D$p$}}$ pair as (\ref{eq:Dp_antiDp_Free}) on the basis of BSFT. For the $p=9$ case, it reads
\BA
  F_o (T, \beta)
    &=& - \ \frac{2 (2 \pi)^4 v_9}{\beta_H^{10}}
      \int_{0}^{\infty} \frac{d \tau}{\tau^6} \ 
        e^{-4 \pi |T|^2 \tau}
          \nonumber \\
  && \hspace{20mm}
    \times \left[ \left\{ \frac{\vartheta_3 (0 | i \tau)}
      {{\vartheta_1}' (0 | i \tau)} \right\}^4
        \left\{ \sum_{w=1}^{\infty}
          \exp \left( - \frac{\pi w^2 \beta^2}{\beta_H^2 \tau}
            \right) \right\}
              \right. \nonumber \\
  && \hspace{30mm} \left.
    - \left\{ \frac{\vartheta_2 (0 | i \tau)}
      {{\vartheta_1}' (0 | i \tau)} \right\}^4
        \left\{ \sum_{w=1}^{\infty} (-1)^w
          \exp \left( - \frac{\pi w^2 \beta^2}{\beta_H^2 \tau}
            \right) \right\} \right].
\label{eq:freeD9antiD9}
\EA
For the later convenience we show the concrete derivation of the cylinder amplitude and the free energy by using the operator formalism. The cylinder amplitude is given by
\BA
  A_{C_2} (T, \beta) &=& 2 \int_0^{\infty} \frac{d \tau}{2 \tau} \ 
    e^{- I_{\partial \Sigma}}
      \left[ \textrm{Tr} \left\{ (-1)^{G^{bc}} b_0 c_0
        (-1)^{G_{NS}^{\beta \gamma}}
          \exp \left( - 2 \pi \tau H_{NS}^{\Sigma} \right) \right\}
            \right. \nonumber \\
  && \hspace{30mm} \left.
    + \textrm{Tr} \left\{ (-1)^{G^{bc}} b_0 c_0
      (-1)^{G_R^{\beta \gamma}}
        \exp \left( - 2 \pi \tau H_R^{\Sigma} \right) \right\}
          \right].
\label{eq:CylAmp}
\EA
The overall factor two comes from the following fact. Let (D9,$\overline{\textrm{D9}}$) represent an open string whose one end is on D9-brane while the other end on $\overline{\textrm{D9}}$-brane. Then we must impose the GSO projection for (D9,D9) and ($\overline{\textrm{D9}}$,$\overline{\textrm{D9}}$), and the opposite GSO projection for (D9,$\overline{\textrm{D9}}$) and ($\overline{\textrm{D9}}$,D9). Thus, the sum of these contributions gives factor two. We choose the form of the boundary action as (\ref{eq:BoundaryAction}), which can be calculated as
\BA
  I_{\partial \Sigma} = 4 \pi \tau |T|^2.
\EA
The Hamiltonian for world-sheet bulk can be expressed as
\BA
  H_{NS}^{\Sigma} &=& \alpha' \sum_{\mu = 0}^9 ({\hat{k}}_{\mu})^2
    + \sum_{\mu = 0}^9 \sum_{l=1}^{\infty}
      \alpha_{-l}^{\mu} \alpha_l^{\mu}
        + \sum_{\mu = 0}^9 \sum_{r = \frac{1}{2}}^{\infty}
          r b_{-r}^{\mu} b_r^{\mu}
            \nonumber \\
  && \hspace{10mm}
    + \sum_{l=1}^{\infty} l \left( b_{-l} c_l + c_{-l} b_l \right)
      + \sum_{r = \frac{1}{2}}^{\infty}
        r \left( \beta_{-r} \gamma_r + \gamma_{-r} \beta_r \right)
          - \ \frac{1}{2},
\EA
for NS sector and
\BA
  H_R^{\Sigma} &=& \alpha' \sum_{\mu = 0}^9 ({\hat{k}}_{\mu})^2
    + \sum_{\mu = 0}^9 \sum_{l=1}^{\infty}
      \alpha_{-l}^{\mu} \alpha_l^{\mu}
        + \sum_{\mu = 0}^9 \sum_{m=1}^{\infty}
          m d_{-m}^{\mu} d_m^{\mu}
            \nonumber \\
  && \hspace{10mm}
    + \sum_{l=1}^{\infty} l \left( b_{-l} c_l + c_{-l} b_l \right)
      + \sum_{m=1}^{\infty}
        m \left( \beta_{-m} \gamma_m + \gamma_{-m} \beta_m \right),
\EA
for R sector. ${\hat{k}}_{\mu}$ denotes momentum operator. $\alpha_n$, $b_r$ and $d_m$ stand for the oscillators of world-sheet bosons, NS fermions and R fermions, respectively, and $c_n$, $b_n$, $\beta_r$ and $\gamma_r$ that of ghosts, anti-ghosts, superconformal ghosts and superconformal anti-ghosts, respectively. The ghost number operator is defined as
\BA
  G^{bc} = \sum_{l=1}^{\infty} \left( c_{-l} b_l - b_{-l} c_l \right)
    + c_0 b_0,
\EA
and the superconformal ghost number operators as
\BA
  G_{NS}^{\beta \gamma}
    &=& - \sum_{r = \frac{1}{2}}^{\infty}
      \left( \gamma_{-r} \beta_r + \beta_{-r} \gamma_r \right), \\
  G_R^{\beta \gamma}
    &=& - \sum_{m=1}^{\infty}
      \left( \gamma_{-m} \beta_m + \beta_{-m} \gamma_m \right)
        - \gamma_0 \beta_0,
\EA
for NS sector and R sector, respectively.

Each part of the trace is calculated as follows. The momentum operator ${\hat{k}}^0$ has discrete eigenvalue, since the Euclidean time is compactified with radius $\beta / (2 \pi)$. In the case of periodic boundary condition in this direction, then the trace over the momentum mode states is calculated as\footnote{We can restrict $k_0$ to be positive and replace the integration measure $d \tau / (2 \tau)$ by $d \tau / \tau$ from the starting point. However, we need to be careful to deal with the difference between $n=0$ term and $w=0$ one in such a case.}
\BA
  \textrm{Tr} \ 
    \exp \left[ - 2 \pi \tau \alpha' ({\hat{k}}_0)^2 \right]
      &=& \sum_{n = - \infty}^{\infty}
        \exp \left[ - 2 \pi \tau \alpha'
          \left( \frac{2 \pi n}{\beta} \right)^2 \right]
            \nonumber \\
  &=& \sum_{w = - \infty}^{\infty}
    \int_{- \infty}^{\infty} du \ \exp (2 \pi i w u)
      \exp \left( - \ \frac{\pi \beta_H^2 \tau}{\beta^2} \ u^2
        \right)
          \nonumber \\
  &=& \frac{\beta}{\beta_H} \ \tau^{- \frac{1}{2}}
    \sum_{w = - \infty}^{\infty}
      \exp \left( - \ \frac{\pi w^2 \beta^2}{\beta_H^2 \tau} \right),
\EA
where we have used the Poisson resummation formula (\ref{eq:PoissonResum}). In the case of antiperiodic boundary condition, we obtain
\BA
  \textrm{Tr} \ 
    \exp \left[ - 2 \pi \tau \alpha' ({\hat{k}}_0)^2 \right]
      = \frac{\beta}{\beta_H} \ \tau^{- \frac{1}{2}}
        \sum_{w = - \infty}^{\infty} (-1)^w
          \exp \left( - \ \frac{\pi w^2 \beta^2}{\beta_H^2 \tau}
            \right).
\EA
In other directions, the trace over momentum mode states gives Gaussian Integral as
\BA
  \textrm{Tr} \ 
    \exp \left[ - 2 \pi \tau \alpha' \sum_{i=1}^9 ({\hat{k}}^i)^2 \right]
      &=& \frac{v_9}{(2 \pi)^9}
        \int_{- \infty}^{\infty} d^9 k \ 
          \exp \left[ - 2 \pi \tau \alpha'
            \sum_{i=1}^9 ({\hat{k}}^i)^2 \right]
              \nonumber \\
  &=& v_9 \beta_H^{-9} \tau^{- \frac{9}{2}}.
\EA
For the oscillators of world-sheet bosons we get
\BA
  \textrm{Tr} \ 
    \exp \left( - 2 \pi \tau
      \sum_{\mu = 0}^9 \sum_{l=1}^{\infty}
        \alpha_{-l}^{\mu} \alpha_l^{\mu} \right)
          = \left[ \prod_{l=1}^{\infty}
            \left\{ 1 - \exp \left( - 2 \pi \tau l \right) \right\}
              \right]^{-10}.
\EA
The contribution from the world-sheet fermions is given by
\BA
  \textrm{Tr} \ 
    \exp \left( - 2 \pi \tau
      \sum_{\mu = 0}^9 \sum_{r = \frac{1}{2}}^{\infty}
        r b_{-r}^{\mu} b_r^{\mu} \right)
          = \left[ \prod_{r = \frac{1}{2}}^{\infty}
            \left\{ 1 + \exp \left( - 2 \pi \tau r \right) \right\}
              \right]^{10},
\EA
for NS sector and
\BA
  \textrm{Tr} \ 
    \exp \left( - 2 \pi \tau
      \sum_{\mu = 0}^9 \sum_{m=1}^{\infty}
        m d_{-m}^{\mu} d_m^{\mu} \right)
          = 2^5 \left[ \prod_{m=1}^{\infty}
            \left\{ 1 + \exp \left( - 2 \pi \tau m \right) \right\}
              \right]^{10},
\label{eq:TrRosc}
\EA
for R sector, respectively. The factor $2^5$ in R sector comes from the sum over zero mode states. The ghost part can be calculated by using the fact that the expectation value of $c_0 b_0$ in the ground states satisfies
\BA
  \langle \downarrow | (-1)^{c_0 b_0}
    b_0 c_0 | \downarrow \rangle &=& 1, \\
  \langle \uparrow | (-1)^{c_0 b_0}
    b_0 c_0 | \uparrow \rangle &=& 0.
\EA
The result is
\BA
  \textrm{Tr} \ 
    \left\{ (-1)^{G_{bc}} b_0 c_0
      \exp \left[ - 2 \pi \tau \sum_{l=1}^{\infty}
        l \left( b_{-l} c_l + c_{-l} b_l \right) \right] \right\}
          = \left[ \prod_{l=1}^{\infty}
            \left\{ 1 - \exp \left( - 2 \pi \tau l \right) \right\}
              \right]^2.
\EA
The part of superconformal ghost is calculated as
\BA
  \textrm{Tr} \left\{ (-1)^{G_{NS}^{\beta \gamma}}
    \exp \left[ - 2 \pi \tau \sum_{r = \frac{1}{2}}^{\infty}
      r \left( \beta_{-r} \gamma_r + \gamma_{-r} \beta_r \right)
        \right] \right\}
          = \left[ \prod_{r = \frac{1}{2}}^{\infty}
            \left\{ 1 + \exp \left( - 2 \pi \tau r \right) \right\}
              \right]^{-2},
                \nonumber\\
\EA
for NS sector, and
\BA
  \textrm{Tr} \left\{ (-1)^{G_R^{\beta \gamma}}
    \exp \left[ - 2 \pi \tau \sum_{m=1}^{\infty}
      m \left( \beta_{-m} \gamma_m + \gamma_{-m} \beta_m \right)
        \right] \right\}
          = \frac{1}{2} \left[ \prod_{m=1}^{\infty}
            \left\{ 1 + \exp \left( - 2 \pi \tau m \right) \right\}
              \right]^{-2},
                \nonumber \\
\EA
for R sector, respectively. The overall factor $1/2$ in the R sector case comes from the infinite sum related to the zero modes. Collecting all these together, we obtain
\BA
  A_{C_2} (T, \beta)
    &=& \frac{(2 \pi)^4 \beta v_9}{\beta_H^{10}}
      \int_0^{\infty} \frac{d \tau}{\tau^6} \ 
        e^{- 4 \pi \tau |T|^2}
          \nonumber \\
  && \hspace{15mm}
    \times \left[ \left\{ \frac{\vartheta_3 \left( 0 | i \tau \right)}
      {\vartheta_1 ' \left( 0 | i \tau \right)} \right\}^4
        \left\{ \sum_{w = - \infty}^{\infty}
          \exp \left( - \ \frac{\pi w^2 \beta^2}{\beta_H^2 \tau}
            \right) \right\}
              \right. \nonumber \\
  && \hspace{20mm} \left.
    - \left\{ \frac{\vartheta_2 \left( 0 | i \tau \right)}
      {\vartheta_1 ' \left( 0 | i \tau \right)} \right\}^4
        \left\{ \sum_{w = - \infty}^{\infty} (-1)^w
          \exp \left( - \ \frac{\pi w^2 \beta^2}{\beta_H^2 \tau}
            \right) \right\} \right].
\EA
The free energy (\ref{eq:freeD9antiD9}) can be obtained by dividing by $(- \beta)$ and omit the $w=0$ term. If we include this $w=0$ term to the free energy, it is interpreted as the vacuum energy, since it does not depend on $\beta$. Thus, we subtract this term.

Now, we are ready to investigate the amplitude in the closed string vacuum limit. For large but finite $|T|$, the contribution of the amplitude comes from only small $\tau$ part. From the boundary action (\ref{eq:BoundaryAction}) we can see that this part corresponds to the tiny hole case. It also corresponds to the high temperature case, since only small $\tau$ part contribute when $\beta$ is close to $\beta_H$. It is convenient to define
\BA
  t = \frac{1}{\tau},
\label{eq:taut}
\EA
and take large $t$ limit. By using the parameter $t$, the amplitude can be rewritten as
\BA
  A_{C_2} (T, \beta)
    &=& \frac{(2 \pi)^4 \beta v_9}{\beta_H^{10}}
      \int_{0}^{\infty}
        dt \ \exp \left( - \ \frac{4 \pi |T|^2}{t} \right)
          \nonumber \\
  && \hspace{15mm}
    \times \left[ \left\{ \frac{\vartheta_3 (0 | i t)}
      {{\vartheta_1}' (0 | it)} \right\}^4
        \left\{ \sum_{w = - \infty}^{\infty}
          \exp \left( - \frac{\pi w^2 \beta^2 t}{\beta_H^2} \right)
            \right\}
              \right. \nonumber \\
  && \hspace{20mm} \left.
    - \left\{ \frac{\vartheta_4 (0 | it)}
      {{\vartheta_1}' (0 | it)} \right\}^4
        \left\{ \sum_{w = - \infty}^{\infty} (-1)^w
          \exp \left( - \frac{\pi w^2 \beta^2 t}{\beta_H^2} \right)
            \right\} \right],
\label{eq:Cylt}
\EA
where we have used the modular transformation of the Jacobi theta functions (\ref{eq:thetaS1}) $\sim$ (\ref{eq:thetaS4}). If we ignore the $w=0$ term, then $w = \pm 1$ terms give the largest contribution for large $t$. The leading term in the large $t$ region is given by
\BE
  A_{C_2} (T, \beta)
    \simeq \frac{4 \beta v_9}{\beta_H^{10}}
      \int_0^{\infty} dt \ 
        \exp \left( - \ \frac{4 \pi |T|^2}{t}
          - \pi \ \frac{\beta^2 - \beta_H^2}
            {\beta_H^2} \ t \right).
\label{eq:FHag}
\EE
In the case of $t \gg |T|^2$, we can ignore the $|T|^2$ term in the exponent in the integrand. Then this leading term of the amplitude corresponds to the propagator of zero momentum winding tachyon. Momentum conservation in the directions with Neumann boundary conditions sets the momentum to zero \cite{VM}. At first glance, it seems that the mass in this case is a half of winding tachyon one (\ref{eq:windtacmass}). However, this difference comes from the factor $2$ difference of (\ref{eq:closedstau2}) and (\ref{eq:openstau}). We can derive exact winding tachyon propagator by replacing $t$ by $2t$. It is noteworthy that above leading term comes from the cylinder world-sheet wrapping once around the compactified Euclidean time. As we have mentioned in \S \ref{sec:HagTra}, the winding tachyon corresponds to the closed string winding once around the compactified Euclidean time.
\begin{figure}[tbp]
\centering
\includegraphics[width=.5\textwidth]{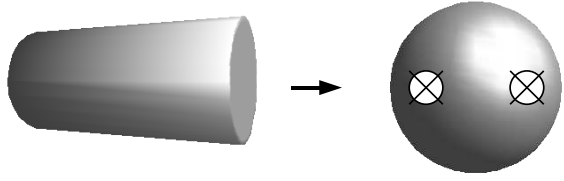}
\caption{\label{fig:Cyl_Sph2pt} The cylinder world-sheet (left) approaches to the sphere world-sheet with two winding tachyons insertion (right).}
\end{figure}

Let us compare this one-loop amplitude in the open string case with the sphere amplitude with two winding tachyons insertion in the closed string case. For the latter case, winding tachyon vertex operators become marginal when $\beta = \beta_H$, and winding tachyon becomes massless. The sphere amplitude for two closed strings vanishes, since it is divided by an infinite volume of residual conformal Killing group \cite{Polchinski}. For the former case, we consider the closed string vacuum limit $|T| \rightarrow \infty$ together with the Hagedorn temperature limit $\beta \rightarrow \beta_H$. By using the modified Bessel function $K_{\nu} (\chi)$, the amplitude can be rewritten as
\BE
  A_{C_2} (T, \beta)
    \simeq \frac{64 \pi v_9 |T|^2}
      {\beta_H^9 \chi} \ K_1 (\chi),
\EE
where we have defined
\BE
  \chi = \frac{4 \pi |T|
    \sqrt{\beta^2 - \beta_H^2}}
      {\beta_H}.
\label{eq:definitionz}
\EE
Let us assume that $\chi \rightarrow \infty$, as $|T| \rightarrow \infty$ and $\beta \rightarrow \beta_H$.\footnote{If we instead take the $\chi \rightarrow 0$ limit, the amplitude diverges.} Then using the asymptotic form of the modified Bessel function
\BE
  K_{\nu} (\chi) \simeq \sqrt{\frac{\pi}{2 \chi}} \ e^{- \chi},
\label{eq:Kzasym}
\EE
we obtain the following form of the amplitude
\BE
  A_{C_2} (T, \beta)
    \simeq \frac{4 \sqrt{2} \ v_9 |T|^{\frac{1}{2}}}
      {\beta_H^{\frac{15}{2}}
        (\beta^2 - \beta_H^2)^{\frac{3}{4}}} \ 
          \exp \left( - \ \frac{4 \pi |T|
            \sqrt{\beta^2 - \beta_H^2}}
              {\beta_H} \right).
\label{eq:FTinfty}
\EE
Thus, if we assume that
\BE
  \frac{|T|^{\frac{1}{2}}}
    {(\beta^2 - \beta_H^2)^{\frac{3}{4}}} \ 
      \exp \left( - \ \frac{4 \pi |T|
        \sqrt{\beta^2 - {\beta_H}^2}}
          {\beta_H} \right)
            \rightarrow 0,
\label{eq:limit1}
\EE
as $|T| \rightarrow \infty$ and $\beta \rightarrow \beta_H$, the one-loop amplitude vanishes. This limit is consistent with $\chi \rightarrow \infty$. Therefore, if we take the appropriate limit, the cylinder amplitude approaches to the sphere amplitude for two winding tachyons, as is depicted in Figure \ref{fig:Cyl_Sph2pt}.

\subsection{Cylinder World-sheet with a Single Massless Boson Insertion}
\label{sec:3vertices}

In this subsection, we show that the cylinder amplitude with a single massless boson insertion approaches to the sphere amplitude with two winding tachyons and a massless boson insertion in the closed string vacuum limit together with the Hagedorn temperature limit. Let us first compute the sphere amplitude for two winding tachyons and a massless boson by using operator product expansion (OPE). This type of amplitude was calculated in the heterotic string case by Schulgin and Troost \cite{heteroamplitude}. We calculate it in type II string theory. In the case of $\beta = \beta_H$, winding tachyon is massless, and its vertex operator is marginal. The vertex operator of zero momentum winding tachyon in $(-1,-1)$ picture is given by \cite{Polchinski}
\BA
  {\cal V}_{w = \pm 1}^{-1,-1} = g_s e^{- \phi - \tilde{\phi}}
    : \exp \left[ \pm i \sqrt{\frac{2}{\alpha '}} \ {\tilde{X}}^0 (z)
      \right] :,
\EA
where we have defined
\BA
  {\tilde{X}}^0 (z)
    = X_L^0 (z) - X_R^0 (\overline{z}),
\EA
and $w$ denotes the winding number. $X_L^0$ describes left-moving modes of $X^0$, and $X_R^0$ right-moving modes. $\phi$ and $\tilde{\phi}$ stand for the bosonized superconformal ghosts. We pick up zero momentum winding tachyon because it is argued that the Hagedorn transition occurs via its condensation. It is consistent to its correspondence to the Neumann boundary condition of open strings on a D9-$\overline{\textrm{D9}}$ pair in the closed string vacuum limit. In order to obtain non-zero amplitude, we should insert the same number of $w=+1$ vertex operators and $w=-1$ ones into the sphere world-sheet. This can be interpreted as the conservation of winding number. We here consider the insertion of a single $w=+1$ vertex operator and a single $w=-1$ one, so that the world-sheet wraps once around the compactified Euclidean time. The vertex operator for NS-NS zero momentum massless boson in $(0,0)$ picture is given by
\BA
  {\cal V}_{e_{\mu \nu}}^{0,0}  = \frac{2 g_s}{\alpha '} \ e_{\mu \nu}
    : \partial X_L^{\mu} (z)
      \overline{\partial} X_R^{\nu} (\overline{z}) :.
\label{eq:vertexmasslessboson}
\EA
Only $\mu =0$, $\nu = 0$ component remains after taking the contraction with the vertex operators of winding tachyon. Massless modes of closed strings in NS-NS sector contain graviton, dilaton and Kalb-Ramond fields. We will insert zero momentum vertex operators into the sphere world-sheet in order that the momentum conservation will be satisfied. We need to evaluate OPE of these vertex operators. The OPE of $X_L$ and $X_R$ is given by
\BA
  X_L^{\mu} (z_1) X_L^{\nu} (z_2)
    &\sim& - \ \frac{\alpha '}{2} \ \delta^{\mu \nu} \ln z_{12}, \\
  X_R^{\mu} ({\overline{z}}_1) X_R^{\nu} ({\overline{z}}_2)
    &\sim& - \ \frac{\alpha '}{2} \ \delta^{\mu \nu}
      \ln {\overline{z}}_{12}, \\
  X_L^{\mu} (z_1) X_R^{\nu} ({\overline{z}}_2)
    &\sim& 0,
\EA
where $z_{ij} = z_i - z_j$. Then the world-sheet boson part of the OPE in the product of above three vertex operators is evaluated as
\BA
  : \exp \left[ i \sqrt{\frac{2}{\alpha '}} \ {\tilde{X}}^0 (z_1)
    \right] :
      : \exp \left[ - i \sqrt{\frac{2}{\alpha '}} \ {\tilde{X}}^0 (z_2)
        \right] :
          : \partial X_L^0 \overline{\partial} X_R^0 (z_3) : \ 
            \sim \frac{\alpha '}{2} \ \frac{1}{|z_{13}|^2 |z_{23}|^2},
\EA
and the expectation value of this part as
\BA
  && \left\langle
    : \exp \left[ i \sqrt{\frac{2}{\alpha '}} \ {\tilde{X}}^0 (z_1)
      \right] :
        : \exp \left[ - i \sqrt{\frac{2}{\alpha '}} \ 
          {\tilde{X}}^0 (z_2) \right] :
            : \partial X_L^0 \overline{\partial} X_R^0 (z_3) :
              \right\rangle_{S_2} \nonumber \\
  && \hspace{80mm} \sim \frac{\alpha '}{2} \ (2 \pi)^{10} C_{S_2}^X \ 
    \frac{1}{|z_{13}|^2 |z_{23}|^2},
\EA
where $C_{S_2}^X$ is a constant. The expectation values of the ghosts and the bosonized superconformal ghosts are given by
\BA
  \left\langle c (z_1) c (z_2) c (z_3)
    c ({\overline{z}}_1) c ({\overline{z}}_2) c ({\overline{z}}_3)
      \right\rangle_{S_2}
        = C_{S_2}^g |z_{12}|^2 |z_{13}|^2 |z_{23}|^2,
\EA
where $C_{S_2}^g$ is also a constant, and
\BA
  \left\langle e^{- \phi (z_1)} e^{- \phi (z_2)} \right\rangle_{S_2}
    &=& \frac{1}{z_{12}}, \\
  \left\langle e^{- \tilde{\phi} ({\overline{z}}_1)}
    e^{- \tilde{\phi} ({\overline{z}}_2)} \right\rangle_{S_2}
    &=& \frac{1}{{\overline{z}}_{12}},
\EA
respectively. Collecting all these together, we obtain the following sphere amplitude
\BA
  A_{S_2} \left( w = +1 ; w = -1 ; k = 0, e_{00} \right)
    &=& e^{- 2 \lambda}
      \left\langle c \tilde{c} {\cal V}_{w = +1}^{-1,-1} (z_1) \ 
        c \tilde{c} {\cal V}_{w = -1}^{-1,-1} (z_2) \ 
          c \tilde{c} {\cal V}_{e_{00}}^{0,0} (z_3) \right\rangle_{S_2}
            \nonumber \\
  &=& 8 \pi (2 \pi)^{10} g_s e_{00},
\label{eq:sphere3pt}
\EA
where we have defined the overall constant as\footnote{Instead of the definition of this constant in Reference \cite{Polchinski}, we have adjusted it such that $A_{S_2}$ is dimensionless.}
\BA
  C_{S_2} = e^{- 2 \lambda} C_{S_2}^X C_{S_2}^g
    = \frac{8 \pi}{g_s^2}.
\EA
In this case the amplitude is non-zero in contrast to the amplitude for two winding tachyons in the previous subsection.

\begin{figure}[tbp]
\centering
\includegraphics[width=.45\textwidth]{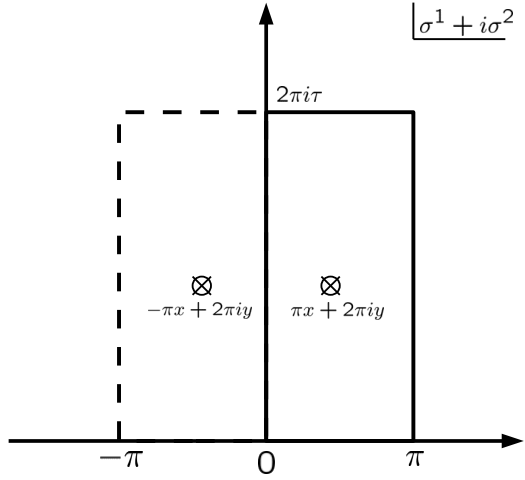}
\caption{\label{fig:DoublingTrick} These are the positions of the vertices on the open string world-sheet when we use the doubling trick.}
\end{figure}
Let us next compute the cylinder amplitude with a single massless boson insertion. We use the doubling trick, since we need to insert the closed string vertex operator on the open string world-sheet. We adopt the world-sheet coordinates $\sigma^{\alpha}$ that are related to the complex coordinate $z$ as
\BA
  z = e^{-i (\sigma^1 + i \sigma^2)},
\EA
where $\sigma^2$ is Euclidean time on world-sheet. We fix the position of the vertex operator at
\BA
  \sigma^1 &=& \pi x, \\
  \sigma^2 &=& 2 \pi y,
\label{eq:ptVer}
\EA
as is depicted in Figure \ref{fig:DoublingTrick}. The ranges of the parameters $x$ and $y$ are given by
\BA
  && 0 < x < 1, \\
  && 0 \leq y < \tau.
\EA
According to the doubling trick, the position of anti-holomorphic part of vertex operator is given by
\BA
  \sigma^1 &=& - \pi x, \\
  \sigma^2 &=& 2 \pi y,
\EA
and anti-holomorphic operators are replaced by holomorphic ones. Then the closed string vertex operator (\ref{eq:vertexmasslessboson}) is replaced by
\BA
  {\cal V}_{e_{\mu \nu}, k^{\mu} = 0}^{0,0} (x,y)
    &=& \frac{2 g_s}{\alpha'} \ e_{\mu \nu}
      : \partial X_L^{\mu} (\pi x + 2 \pi i y) :
        : \partial X_L^{\nu} (- \pi x + 2 \pi i y) :
          \nonumber \\
  &=& - g_s e_{\mu \nu} \sum_{l_1 , l_2 = - \infty}^{\infty}
    \alpha_{l_1}^{\mu} \alpha_{l_2}^{\nu}
      \exp \left[ \pi i l_1 (x+2iy) \right]
        \exp \left[ - \pi i l_2 (x-2iy) \right],
\label{eq:vertexmasslessbosondoubling}
\EA
where $\alpha_0^{\mu}$ is defined as
\BA
  \alpha_0^{\mu} = (2 \alpha')^{\frac{1}{2}} {\hat{k}}^{\mu}.
\EA
Using this vertex operator, we can express the cylinder amplitude with a single massless boson insertion as
\BA
  A_{C_2} (T, \beta)
    &=& 2 \int_0^{\infty} \frac{d \tau}{2 \tau} \ 
      e^{- I_{\partial \Sigma}}
        \left[ \textrm{Tr} \left\{ (-1)^{G^{bc}} b_0 c_0
          (-1)^{G_{NS}^{\beta \gamma}}
            \exp \left( - 2 \pi y H_{NS}^{\Sigma} \right)
              \right. \right. \nonumber \\
  && \hspace{40mm} \left.
    \times {\cal V}_{e_{\mu \nu}, k^{\mu} = 0}^{0,0} (x,y)
      \exp \left( - 2 \pi (\tau - y) H_{NS}^{\Sigma} \right)
        \right\}
          \nonumber \\
  && \hspace{30mm} \left.
    - \textrm{Tr} \left\{ (-1)^{G^{bc}} b_0 c_0
      (-1)^{G_R^{\beta \gamma}}
        \exp \left( - 2 \pi \tau H_R^{\Sigma} \right)
          \right. \right. \nonumber \\
  && \hspace{40mm} \left. \left.
    \times {\cal V}_{e_{\mu \nu}, k^{\mu} = 0}^{0,0} (x,y)
      \exp \left( - 2 \pi (\tau - y) H_R^{\Sigma} \right)
        \right\} \right].
\label{eq:CylAmpMasslessVer}
\EA
We only need to calculate the world-sheet boson part, since the other parts are the same in the case of the cylinder amplitude without vertex operator insertion. The non-zero contributions come from the $l_1 = - l_2$ terms in the vertex operator (\ref{eq:vertexmasslessbosondoubling}). For the momentum mode in the Euclidean time direction, the trace is calculated as
\BA
  && \hspace*{-10mm}
    \textrm{Tr} \left\{
      \exp \left[ - 2 \pi y \alpha' ({\hat{k}}^0)^2 \right]
        \left[ - g_s e_{00} \left( \alpha_0^0 \right)^2 \right]
          \exp \left[ - 2 \pi (\tau - y) \alpha' ({\hat{k}}^0)^2
            \right] \right\}
              \nonumber \\
  && = - 2 g_s \alpha' e_{00}
    \left( \frac{2 \pi}{\beta} \right)^2
      \sum_{n = - \infty}^{\infty} n^2
        \exp \left[ - 2 \pi \tau \alpha'
          \left( \frac{2 \pi n}{\beta} \right)^2 \right]
            \nonumber \\
  && = - g_s e_{00} \ \frac{\beta_H^2}{\beta^2}
    \left( - \ \frac{\beta^2}{\pi \beta_H^2} \ 
      \frac{\partial}{\partial \tau} \right)
        \left\{ \sum_{n = - \infty}^{\infty}
          \exp \left( - \ \frac{\pi \beta_H^2 \tau}{\beta^2} \ n^2
            \right) \right\}
              \nonumber \\
  && = \frac{g_s e_{00}}{\pi} \ 
    \frac{\partial}{\partial \tau}
      \left\{ \sum_{w = - \infty}^{\infty}
        \int_{- \infty}^{\infty} du \ \exp (2 \pi i w u)
          \exp \left( - \ \frac{\pi \beta_H^2 \tau}{\beta^2} \ u^2
            \right) \right\}
              \nonumber \\
  && = \frac{g_s e_{00}}{\pi} \ \frac{\beta}{\beta_H}
    \left\{ - \ \frac{1}{2 \tau^{\frac{3}{2}}}
      \sum_{w = - \infty}^{\infty}
        \exp \left( - \ \frac{\pi w^2 \beta^2}{\beta_H^2 \tau}
          \right)
            \right. \nonumber \\
  && \hspace{40mm} \left.
    + \frac{\pi \beta^2}{\beta_H^2 \tau^{\frac{5}{2}}}
      \sum_{w = - \infty}^{\infty} w^2
        \exp \left( - \ \frac{\pi w^2 \beta^2}{\beta_H^2 \tau}
          \right) \right\},
\EA
in the case of periodic boundary condition, and
\BA
  && \hspace*{-10mm}
    \textrm{Tr} \left\{
      \exp \left[ - 2 \pi y \alpha' ({\hat{k}}^0)^2 \right]
        \left[ - g_s e_{00} \left( \alpha_0^0 \right)^2 \right]
          \exp \left[ - 2 \pi (\tau - y) \alpha' ({\hat{k}}^0)^2
            \right] \right\} \nonumber \\
  && = \frac{g_s e_{00}}{\pi} \ \frac{\beta}{\beta_H}
    \left\{ - \ \frac{1}{2 \tau^{\frac{3}{2}}}
      \sum_{w = - \infty}^{\infty} (-1)^w
        \exp \left( - \ \frac{\pi w^2 \beta^2}{\beta_H^2 \tau}
          \right)
            \right. \nonumber \\
  && \hspace{40mm} \left.
    + \frac{\pi \beta^2}{\beta_H^2 \tau^{\frac{5}{2}}}
      \sum_{w = - \infty}^{\infty} (-1)^w w^2
        \exp \left( - \ \frac{\pi w^2 \beta^2}{\beta_H^2 \tau}
          \right) \right\},
\EA
in the case of antiperiodic boundary condition, respectively. We have used the Poisson resummation formula (\ref{eq:PoissonResum}). In other directions, the trace over momentum mode states gives
\BA
  \textrm{Tr} \left\{
    \exp \left[ - 2 \pi y \alpha' ({\hat{k}}^i)^2 \right]
      \left[ - g_s e_{ii} \left( \alpha_0^i \right)^2 \right]
        \exp \left[ - 2 \pi (\tau - y) \alpha' ({\hat{k}}^i)^2
          \right] \right\}
            = - \ \frac{g_s e_{ii} L_i}
              {2 \pi \beta_H \tau^{\frac{3}{2}}},
\EA
where $L_i$ is the length we are considering in this direction. For the oscillation modes, we need to do a lengthy calculation as follows. In the case of $l_1 = - l_2 = -l$ with $l>0$, the trace is computed as
\BA
  && \hspace*{-10mm}
    \textrm{Tr} \left\{
      \exp \left[ - 2 \pi y \alpha_{-l}^{\mu} \alpha_l^{\mu} \right]
        \left[ - g_s e_{\mu \mu} \alpha_{-l}^{\mu} \alpha_l^{\mu}
          \exp \left[ - \pi il (x+2iy) \right]
            \exp \left[ - \pi il (x-2iy) \right] \right]
              \right. \nonumber \\
  && \hspace{80mm} \left.
    \times \exp \left[ - 2 \pi (\tau - y)
      \alpha_{-l}^{\mu} \alpha_l^{\mu} \right] \right\}
        \nonumber \\
  && = - g_s e_{\mu \mu} \exp (-2 \pi ilx) l
    \sum_{n_l^{\mu} = 1}^{\infty} n_l^{\mu}
      \exp \left( - 2 \pi l n_l^{\mu} \tau \right)
        \nonumber \\
  && = - g_s e_{\mu \mu} \exp (-2 \pi ilx) l
    \left( - \ \frac{1}{2 \pi l} \ 
      \frac{\partial}{\partial \tau} \right)
        \left\{ \sum_{n_l^{\mu} = 1}^{\infty}
          \exp \left( - 2 \pi l n_l^{\mu} \tau \right) \right\}
            \nonumber \\
  && = - \ \frac{g_s e_{\mu \mu} l
    \exp (-2 \pi ilx) \exp \left( - 2 \pi l \tau \right)}
      {\left\{ 1 - \exp \left( - 2 \pi l \tau \right) \right\}^2}.
\EA
Similarly, in the case of $l_1 = - l_2 = l$ we get
\BA
  && \hspace*{-10mm}
    \textrm{Tr} \left\{
      \exp \left[ - 2 \pi y \alpha_{-l}^{\mu} \alpha_l^{\mu} \right]
        \left[ - g_s e_{\mu \mu} \alpha_l^{\mu} \alpha_{-l}^{\mu}
          \exp \left[ \pi il (x+2iy) \right]
            \exp \left[ \pi il (x-2iy) \right] \right]
              \right. \nonumber \\
  && \hspace{80mm} \left.
    \times \exp \left[ - 2 \pi (\tau - y)
      \alpha_{-l}^{\mu} \alpha_l^{\mu} \right] \right\}
        \nonumber \\
  && = - \ \frac{g_s e_{\mu \mu} l \exp (2 \pi ilx)}
    {\left\{ 1 - \exp \left( - 2 \pi l \tau \right)
      \right\}^2}.
\EA
For the oscillation modes in total we obtain
\BA
  && \hspace*{-10mm}
    \textrm{Tr} \left\{
      \exp \left[ - 2 \pi y \sum_{\mu_3 = 0}^9 \sum_{l_3 = 1}^{\infty}
        \alpha_{- l_3}^{\mu_3} \alpha_{l_3}^{\mu_3} \right]
          \right. \nonumber \\
  && \times \left[ - g_s e_{\mu \nu}
    {\sum_{l_1 , l_2 = - \infty}^{\infty}}'
      \alpha_{l_1}^{\mu} \alpha_{l_2}^{\nu}
        \exp \left[ \pi i l_1 (x+2iy) \right]
          \exp \left[ - \pi i l_2 (x-2iy) \right] \right]
            \nonumber \\
  && \hspace{50mm} \left.
    \times \exp \left[ - 2 \pi (\tau - y)
      \sum_{\mu_4 = 0}^9 \sum_{l_4 = 1}^{\infty}
        \alpha_{- l_4}^{\mu_4} \alpha_{l_4}^{\mu_4} \right]
          \right\}
            \nonumber \\
  && = - g_s \left[ \prod_{l=1}^{\infty}
    \left\{ 1 - \exp \left( - 2 \pi \tau l \right) \right\}
      \right]^{-10}
        \nonumber \\
  && \hspace{10mm}
    \times \sum_{\mu = 0}^9 e_{\mu \mu}
      \sum_{l=1}^{\infty}
        \frac{l \exp (-2 \pi ilx)
          \exp \left( - 2 \pi l \tau \right)
            + l \exp (2 \pi ilx)}
              {1 - \exp \left( - 2 \pi l \tau \right)},
\label{eq:traceosctotal}
\EA
where the prime over the sum indicates the exclusion of $l_1 = 0$ and $l_2 = 0$. In order to use the identity
\BA
  \sum_{l=1}^{\infty} \frac{1}{l} \ \frac{u^l}{1 - v^l}
    = - \sum_{n=0}^{\infty} \ln \left( 1 - u v^n \right).
\EA
we rewrite the sum over $l$ as
\BA
  && \hspace*{-10mm}
    \sum_{l=1}^{\infty}
      \frac{l \exp (-2 \pi ilx) \exp \left( - 2 \pi l \tau \right)
        + l \exp (2 \pi ilx)}
          {1 - \exp \left( - 2 \pi l \tau \right)}
            \nonumber \\
  && = - \ \frac{1}{4 \pi^2} \ \frac{\partial^2}{\partial x^2}
    \left[ \sum_{l=1}^{\infty}
      \left[ \frac{\exp (-2 \pi ilx)
        \exp \left( - 2 \pi l \tau \right)}
          {l \left\{ 1 - \exp \left( - 2 \pi l \tau \right) \right\}}
            \right. \right. \nonumber \\
  && \hspace{50mm} \left. \left.
    + \frac{\exp (2 \pi ilx)}
      {l \left\{ 1 - \exp \left( - 2 \pi l \tau \right) \right\}}
        + f (l , \tau) \right] \right],
\EA
where $f (l , \tau)$ is an arbitrary function of $l$ and $\tau$. We choose $f (l , \tau)$ as
\BA
  f (l , \tau)
    = \frac{- 2 \exp \left( - 2 \pi l \tau \right)}
      {l \left\{ 1 - \exp \left( - 2 \pi l \tau \right) \right\}},
\EA
so that the terms in the sum over $l$ are combined to be the logarithm of the theta functions as follows:
\BA
  && \hspace*{-10mm}
    \sum_{l=1}^{\infty}
      \left[ \frac{\exp (-2 \pi ilx) \exp (- 2 \pi l \tau)}
        {l \left\{ 1 - \exp (- 2 \pi l \tau) \right\}}
          + \frac{\exp (2 \pi ilx)}
            {l \left\{ 1 - \exp (- 2 \pi l \tau) \right\}}
              + \frac{- 2 \exp (- 2 \pi l \tau)}
                {l \left\{ 1 - \exp (- 2 \pi l \tau) \right\}}
                  \right]
                    \nonumber \\
  && = \sum_{n=0}^{\infty} \left\{
    \ln \left[ 1
      - \exp (-2 \pi ix)
        \exp \left[ - 2 \pi (n+1) \tau \right] \right]
          + \ln \left[ 1 - \exp (2 \pi ix) \exp (- 2 \pi n \tau)
            \right]
              \right. \nonumber \\
  && \hspace{90mm} \left.
    - 2 \ln \left[ 1 - \exp \left[ - 2 \pi (n+1) \tau \right] \right]
      \right\}
        \nonumber \\
  && = \ln \left[ 1 - \exp (2ix) \right]
    \nonumber \\
  && \hspace{10mm}
    + \sum_{n=1}^{\infty} \ln \left[
      \frac{\left\{ 1 - \exp (2 \pi ix) \exp (- 2 \pi n \tau) \right\}
        \left\{ 1 - \exp (-2 \pi ix) \exp (- 2 \pi n \tau) \right\}}
          {\left\{ 1 - \exp (- 2 \pi n \tau) \right\}^2} \right]
            \nonumber \\
  && = \ln \left[ -2i \exp (\pi ix) \sin (\pi x)
    \prod_{n=1}^{\infty}
      \frac{1 - 2 \cos (2 \pi x) \exp (- 2 \pi n \tau)
        + \exp (- 4 \pi n \tau)}
          {\left\{ 1 - \exp (- 2 \pi n \tau) \right\}^2} \right]
            \nonumber \\
  && = \ln \left[ -2 \pi i \exp (\pi ix) \ 
    \frac{\vartheta_1 (x | i \tau)}{\vartheta_1 ' (0 | i \tau)}
      \right].
\EA
The sum over $l$ can be expressed neatly as
\BA
  \sum_{l=1}^{\infty}
    \frac{l \exp (-2 \pi ilx) \exp \left( - 2 \pi l \tau \right)
      + l \exp (2 \pi ilx)}
        {1 - \exp \left( - 2 \pi l \tau \right)}
          = - \ \frac{1}{4 \pi^2} \ \frac{\partial^2}{\partial x^2}
            \left\{ \ln \left[ \vartheta_1 (x | i \tau) \right]
              \right\},
\EA
since the second derivative of the factors in the logarithm with respect to $x$ vanishes except for $\vartheta_1 (x | i \tau)$. Substituting this into (\ref{eq:traceosctotal}), we get
\BA
  && \hspace*{-10mm}
    \textrm{Tr} \left\{
      \exp \left[ - 2 \pi y \sum_{\mu_3 = 0}^9 \sum_{l_3 = 1}^{\infty}
        \alpha_{- l_3}^{\mu_3} \alpha_{l_3}^{\mu_3} \right]
          \right. \nonumber \\
  && \times \left[ - g_s e_{\mu \nu}
    {\sum_{l_1 , l_2 = - \infty}^{\infty}}'
      \alpha_{l_1}^{\mu} \alpha_{l_2}^{\nu}
        \exp \left[ \pi i l_1 (x+2iy) \right]
          \exp \left[ - \pi i l_2 (x-2iy) \right] \right]
            \nonumber \\
  && \hspace{50mm} \left.
    \times \exp \left[ - 2 \pi (\tau - y)
      \sum_{\mu_4 = 0}^9 \sum_{l_4 = 1}^{\infty}
        \alpha_{- l_4}^{\mu_4} \alpha_{l_4}^{\mu_4} \right]
          \right\}
            \nonumber \\
  && = \frac{g_s}{4 \pi^2} \left( \sum_{\mu = 0}^9 e_{\mu \mu} \right)
    \left[ \prod_{l=1}^{\infty}
      \left\{ 1 - \exp \left( - 2 \pi \tau l \right) \right\}
        \right]^{-10}
          \frac{\partial^2}{\partial x^2}
            \left\{ \ln \left[ \vartheta_1 (x | i \tau) \right]
              \right\}.
\EA
Now the calculation of all the parts of the trace in (\ref{eq:CylAmpMasslessVer}) are complete. Gathering contribution from all these parts, we obtain
\BA
  A_{C_2} (T, \beta)
    &=& \frac{(2 \pi)^3 g_s \beta v_9}{\beta_H^{10}}
      \int_0^{\infty} \frac{d \tau}{\tau^6} \ 
        e^{- 4 \pi \tau |T|^2}
          \nonumber \\
  && \hspace{10mm}
    \times \left[
      \left\{ \frac{\vartheta_3 \left( 0 | i \tau \right)}
        {\vartheta_1 ' \left( 0 | i \tau \right)} \right\}^4
          \left[ \frac{2 \pi e_{00} \beta^2}{\beta_H^2} \ \tau^{-2}
            \sum_{w = - \infty}^{\infty} w^2
              \exp \left( - \ \frac{\pi w^2 \beta^2}{\beta_H^2 \tau}
                \right)
                  \right. \right. \nonumber \\
  && \hspace{40mm}
    + \left( \sum_{\mu = 0}^9 e_{\mu \mu} \right)
      \left[ \tau^{-1}
        + \frac{1}{2 \pi} \ \frac{\partial^2}{\partial x^2}
          \left\{ \ln \left[ \vartheta_1 (x | i \tau) \right]
            \right\} \right]
              \nonumber \\
  && \hspace{60mm} \left.
    \times \left\{ \sum_{w = - \infty}^{\infty}
      \exp \left( - \ \frac{\pi w^2 \beta^2}{\beta_H^2 \tau}
        \right) \right\} \right]
          \nonumber \\
  && \hspace{10mm}
    - \left\{ \frac{\vartheta_2 \left( 0 | i \tau \right)}
      {\vartheta_1 ' \left( 0 | i \tau \right)} \right\}^4
        \left[ \frac{2 \pi e_{00} \beta^2}{\beta_H^2} \ \tau^{-2}
          \sum_{w = - \infty}^{\infty} (-1)^w w^2
            \exp \left( - \ \frac{\pi w^2 \beta^2}{\beta_H^2 \tau}
              \right)
                \right. \nonumber \\
  && \hspace{40mm}
    + \left( \sum_{\mu = 0}^9 e_{\mu \mu} \right)
      \left[ \tau^{-1}
        + \frac{1}{2 \pi} \ \frac{\partial^2}{\partial x^2}
          \left\{ \ln \left[ \vartheta_1 (x | i \tau) \right]
            \right\} \right]
              \nonumber \\
  && \hspace{70mm} \left. \left.
    \times \left\{ \sum_{w = - \infty}^{\infty} (-1)^w
      \exp \left( - \ \frac{\pi w^2 \beta^2}{\beta_H^2 \tau}
        \right) \right\} \right] \right].
          \nonumber \\
\EA
We can infer this form of amplitude from the scattering amplitude at the one-loop level in string theory (see, e.g., Ref. \cite{GSW}).

\begin{figure}[tbp]
\centering
\includegraphics[width=.45\textwidth]{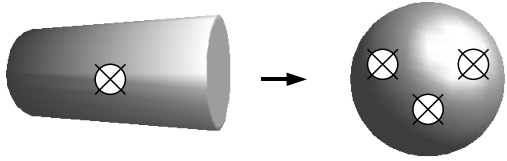}
\caption{\label{fig:Cyl1pt_Sph3pt} The cylinder world-sheet with a massless boson insertion (left) approaches to the sphere world-sheet with two winding tachyons and a massless boson insertion (right).}
\end{figure}

Let us turn to the comparison of this amplitude with the sphere one. We define $t$ as (\ref{eq:taut}) and take large $t$ limit like in the previous subsection. Using the modular transformation of the Jacobi theta functions (\ref{eq:thetaS1}) $\sim$ (\ref{eq:thetaS4}) and
\BA
  \vartheta_1 ( v | \tau )
    = e^{\frac{3 \pi i}{4}} \tau^{- \frac{1}{2}}
      e^{\frac{- \pi i v^2}{\tau}} \vartheta_1
        \left( \frac{v}{\tau} \left| - \frac{1}{\tau} \right. \right),
\EA
we can rewrite the amplitude as
\BA
  A_{C_2} (T, \beta)
    &=& \frac{(2 \pi)^3 g_s \beta v_9}{\beta_H^{10}}
      \int_0^{\infty}dt \ t^4
        \exp \left( - \ \frac{4 \pi |T|^2}{t} \right)
          \nonumber \\
  && \hspace{10mm}
    \times \left[
      \left\{ \frac{\vartheta_3 (0 | it)}
        {\vartheta_1 ' (0 | it)} \right\}^4
          \left[ \frac{2 \pi e_{00} \beta^2}{\beta_H^2} \ t^2
            \sum_{w = - \infty}^{\infty} w^2
              \exp \left( - \ \frac{\pi w^2 \beta^2 t }{\beta_H^2}
                \right)
                  \right. \right. \nonumber \\
  && \hspace{30mm}
    + \left( \sum_{\mu = 0}^9 e_{\mu \mu} \right)
      \left[ t
        + \frac{1}{2 \pi} \ \frac{\partial^2}{\partial x^2}
          \left\{ \ln \left[ - i t^{\frac{1}{2}} e^{- \pi x^2 t}
            \vartheta_1 (ixt | it) \right] \right\} \right]
              \nonumber \\
  && \hspace{70mm} \left.
    \times \left\{ \sum_{w = - \infty}^{\infty}
      \exp \left( - \ \frac{\pi w^2 \beta^2 t}{\beta_H^2}
        \right) \right\} \right]
          \nonumber \\
  && \hspace{10mm}
    - \left\{ \frac{\vartheta_4 (0 | it)}
      {\vartheta_1 ' (0 | it)} \right\}^4
        \left[ \frac{2 \pi e_{00} \beta^2}{\beta_H^2} \ t^2
          \sum_{w = - \infty}^{\infty} (-1)^w w^2
            \exp \left( - \ \frac{\pi w^2 \beta^2 t}{\beta_H^2}
              \right)
                \right. \nonumber \\
  && \hspace{30mm}
    + \left( \sum_{\mu = 0}^9 e_{\mu \mu} \right)
      \left[ t
        + \frac{1}{2 \pi} \ \frac{\partial^2}{\partial x^2}
          \left\{ \ln \left[ - i t^{\frac{1}{2}} e^{- \pi x^2 t}
            \vartheta_1 (ixt | it) \right] \right\} \right]
              \nonumber \\
  && \hspace{60mm} \left. \left.
    \times \left\{ \sum_{w = - \infty}^{\infty} (-1)^w
      \exp \left( - \ \frac{\pi w^2 \beta^2 t}{\beta_H^2}
        \right) \right\} \right] \right].
          \nonumber \\
\EA
Since the part which includes $x$ is calculated as
\BA
  t + \frac{1}{2 \pi} \ \frac{\partial^2}{\partial x^2}
    \left\{ \ln \left[ - i t^{\frac{1}{2}} e^{- \pi x^2 t}
      \vartheta_1 (ixt | it) \right] \right\}
        = \frac{1}{2 \pi} \ \frac{\partial^2}{\partial x^2}
          \left\{ \ln \left[ \vartheta_1 (ixt | it) \right] \right\},
\EA
and the leading terms in $\vartheta_1 (ixt | it)$ is given by
\BA
  \vartheta_1 (ixt | it)
    \simeq i e^{- \left( \frac{1}{4} - x \right) \pi t}
      \left( 1 - e^{- 2 \pi t} \right)
        \left\{ 1 - e^{- 2 \pi (1-x) t} \right\},
\EA
for large $t$, the amplitude is approximated as
\BA
  A_{C_2} (T, \beta)
    &\simeq& \frac{2 g_s v_9}{\beta_H^9}
      \int_0^{\infty}dt \ 
        \exp \left( - \ \frac{4 \pi |T|^2}{t} \right)
          \nonumber \\
  && \hspace{10mm}
    \times \left[ e^{\pi t}
      \left[ e_{00} t^2
        \exp \left( - \ \frac{\pi \beta^2 t }{\beta_H^2} \right)
          \right. \right. \nonumber \\
  && \hspace{30mm} \left.
    - \left( \sum_{\mu = 0}^9 e_{\mu \mu} \right) t^2
      e^{- 2 \pi (1-x) t}
        \exp \left( - \ \frac{\pi \beta^2 t}{\beta_H^2} \right)
          \right]
            \nonumber \\
  && \hspace{20mm}
    + e^{\pi t}
      \left[ e_{00} t^2
        \exp \left( - \ \frac{\pi \beta^2 t}{\beta_H^2} \right)
          \right. \nonumber \\
  && \hspace{30mm} \left. \left.
    - \left( \sum_{\mu = 0}^9 e_{\mu \mu} \right) t^2
      e^{- 2 \pi (1-x) t}
        \exp \left( - \ \frac{\pi \beta^2 t}{\beta_H^2}
          \right) \right] \right],
\EA
where we have omitted the $w=0$ terms. The largest contribution comes from the terms which only include $e_{00}$ because $x<1$. If we keep only this largest contribution, then we obtain
\BA
  A_{C_2} (T, \beta)
    \simeq \frac{4 g_s e_{00} v_9}{\beta_H^9}
      \int_0^{\infty}dt \ t^2
        \exp \left( - \ \frac{4 \pi |T|^2}{t}
          - \pi \ \frac{\beta^2 - \beta_H^2}{\beta_H^2} \ t \right).
\EA
Like in the previous subsection, the leading term in the large $t$ region comes from $w=1$ terms, which correspond to the cylinder world-sheet wrapping once around the compactified Euclidean time. By using the modified Bessel function and its asymptotic form (\ref{eq:Kzasym}), we obtain
\BA
  A_{C_2} (T, \beta)
    \simeq \frac{2^4 \sqrt{2} \ g_s e_{00} v_9 |T|^{\frac{5}{2}}}
      {{\beta_H}^{\frac{11}{2}}
        \left( \beta^2 - {\beta_H}^2 \right)^{\frac{7}{4}}} \ 
          \exp \left( - \ \frac{4 \pi |T|
            \sqrt{\beta^2 - {\beta_H}^2}}
              {\beta_H} \right).
\EA
Finally, if we take $|T| \rightarrow \infty$ and $\beta \rightarrow \beta_H$ limit with following quantity held fixed
\BE
  \frac{|T|^{\frac{7}{2}}} {(\beta^2 - \beta_H^2)^{\frac{7}{4}}} \ 
    \exp \left( - \ \frac{4 \pi |T| \sqrt{\beta^2 - \beta_H^2}}
      {\beta_H} \right)
        = \frac{2^{\frac{17}{2}} \pi^{11} {\beta_H}^{\frac{11}{2}}}
          {v_9},
\label{eq:limit2}
\EE
the amplitude is approximated as
\BA
  A_{C_2} \left( T , \beta ; k=0 , e_{00} \right)
    \simeq 8 \pi (2 \pi)^{10} g_s e_{00},
\EA
which agrees with sphere amplitude (\ref{eq:sphere3pt}) including numerical constant. Therefore, if we take the appropriate limit, the cylinder amplitude with a single massless boson insertion approaches to the sphere amplitude with two winding tachyons and a single massless boson insertion, as is depicted in Figure \ref{fig:Cyl1pt_Sph3pt}. It should be noted that this limit is consistent with (\ref{eq:limit1}) in the previous subsection. In other words, if we take this limit, then (\ref{eq:limit1}) is automatically satisfied.

\section{Stable Minimum near the Hagedorn Temperature}
\label{sec:StableMinimum}

Atick and Witten proposed that the Hagedorn transition occurs via condensation of winding tachyon, and tried to look for the minimum of the potential of winding tachyon \cite{AW}. We have not succeeded in finding it so far. In the previous section, we have shown that this winding tachyon corresponds to the boundary of open string in the closed string vacuum limit $|T| \rightarrow \infty$. In this sense, the proposal of Atick and Witten corresponds to the search of the potential minimum at the closed string vacuum. However, we now aware the space of open string tachyon field. It is reasonable to look for the stable minimum in entire space of the open string tachyon field. We compare the potential energy at closed string vacuum with that at the open string vacuum slightly below the Hagedorn temperature.

Let us first estimate the potential energy at the closed string vacuum. Atick and Witten argued that the Hagedorn transition is a first order phase transition \cite{AW}. This means that the winding tachyon has an expectation value at the potential minimum even under the Hagedorn temperature. The closed string vacuum without winding tachyon condensation is merely a local minimum of the winding tachyon potential in this case. It is preferable that we compare the potential energy at the open string vacuum with that at the minimum of the winding tachyon potential. However, we have not succeeded in computing the winding tachyon potential so far. We evaluate the potential energy at the closed string vacuum without winding tachyon condensation instead. In this case the leading term of the finite temperature effective potential is given by the one-loop free energy of closed string gas. In order to see the concrete value of free energy of closed string gas, it is convenient to rewrite that in the S-representation as
\BE
  F_c (\beta) = - \ \frac{4 (2 \pi)^8 v_9}
    {\beta_H^{10}}
      \int_E \frac{d^2 \tau}{\tau_2^6}
        \left| \frac{\vartheta_2 (0 | \tau)}
          {{\vartheta_1}' (0 | \tau)} \right|^8
            \left[ \sum_{w=1}^{\infty} \left\{ 1 - (-1)^w \right\}
              \exp \left( - \ \frac{2 \pi w^2 \beta^2}
                {\beta_H^2 \tau_2} \right)
                  \right],
\label{eq:Fclosed}
\EE
by using the identity (\ref{eq:aequatioidentica}). Let us define $d_n$ as
\BE
  \sum_{n=0}^{\infty} d_n e^{2 \pi i n \tau}
    \equiv 16 \pi^4 \left\{ \frac{\vartheta_2 (0 | \tau)}
      {{\vartheta_1}' (0 | \tau)} \right\}^4.
\EE
For large $n$, asymptotic formula of $d_n$ is given by (see, e.g., Ref. \cite{GSW})
\BE
  d_n \simeq 2^{- \frac{11}{4}} n^{- \frac{11}{4}}
    \exp \left( 2 \pi \sqrt{2n} \right).
\label{eq:largendn}
\EE
This large $n$ part contribute to the free energy (\ref{eq:Fclosed}) when $\tau_2$ is small. Substituting this asymptotic formula into (\ref{eq:Fclosed}), we obtain
\BA
  F_c (\beta) &\simeq& - \ \frac{4 (2 \pi)^8 v_9}
    {\beta_H^{10}}
      \int_0^{\Lambda} \frac{d^2 \tau}{\tau_2^6}
        \int_{- \frac{1}{2}}^{\frac{1}{2}}
          \times \frac{1}{2^{\frac{11}{2}} (16 \pi^4)^2}
            \nonumber \\
  && \hspace{20mm}
    \times \sum_{n,n'} (n n')^{- \frac{11}{4}}
      \exp \left[ 2 \pi \sqrt{2}
        \left( \sqrt{n} + \sqrt{n'} \right) \right]
          \nonumber \\
  && \hspace{30mm}
    \times \exp \left[ 2 \pi i (n-n') \tau_1
      - 2 \pi (n+n') \tau_2 \right]
        \times 2 \exp \left( - \ 
          \frac{2 \pi \beta^2}{\beta_H^2 \tau_2} \right)
            \nonumber \\
  &=& - \ \frac{v_9}
    {2^{\frac{5}{2}} \beta_H^{10}}
      \int_0^{\Lambda} \frac{d^2 \tau}{\tau_2^6} \ 
        \exp \left( - \ \frac{2 \pi \beta^2}
          {\beta_H^2 \tau_2} \right)
            \left\{ \sum_{n} n^{- \frac{11}{2}}
              \exp \left( 4 \pi \sqrt{2n} - 4 \pi \tau_2 n \right) \right\},
\label{eq:Fclosmalltau}
\EA
where we have introduced the cut off $\Lambda$ in order to guarantee the approximation (\ref{eq:largendn}). In the second equality we have used the fact that the integral of $\tau_1$ yields $n = n'$. If we replace $\tau_2 n$ by $\lambda$ in the case that $\tau_2 \ll 1$, then we can approximate the summation over $n$ by an integral with respect to $\lambda$ as
\BA
  && \sum_{n} n^{- \frac{11}{2}}
    \exp \left( 4 \pi \sqrt{2n} - 4 \pi \tau_2 n \right)
      \nonumber \\
  && \rightarrow {\tau_2}^{\frac{9}{2}}
    \int_0^{\infty} d \lambda \ \lambda^{- \frac{11}{2}}
      \exp \left( 4 \pi \sqrt{\frac{2 \lambda}{\tau_2}}
        - 4 \pi \lambda \right)
          \nonumber \\
  && = 2 {\tau_2}^{\frac{9}{2}} e^{\frac{2 \pi}{\tau_2}}
    \int_{- \frac{1}{\sqrt{2 \tau_2}}}^{\infty} d \rho \ 
      \left( \rho + \frac{1}{\sqrt{2 \tau_2}} \right)^{-10}
        e^{- 4 \pi \rho^2}
          \nonumber \\
  && \simeq 2 {\tau_2}^{\frac{9}{2}} e^{\frac{2 \pi}{\tau_2}}
    \int_{- \infty}^{\infty} d \rho \ 
      \left( \frac{1}{\sqrt{2 \tau_2}} \right)^{-10}
        e^{- 4 \pi \rho^2}
          \nonumber \\
  && = 2^5 {\tau_2}^{\frac{19}{2}} e^{\frac{2 \pi}{\tau_2}},
\EA
where we have defined $\rho$ as
\BA
  \rho = \lambda^{\frac{1}{2}} - \ \frac{1}{\sqrt{2 \tau_2}}.
\EA
Substituting this into (\ref{eq:Fclosmalltau}) and setting the $\beta \simeq \beta_H$, we obtain
\BA
  F_c (\beta) &\simeq& - \ \frac{2^{\frac{5}{2}} v_9}{\beta_H^{10}}
    \int_0^{\Lambda} d \tau_2 \ \tau_2^{\frac{7}{2}} \ 
      \exp \left[ - \ \frac{2 \pi}{\beta_H^2 \tau_2}
        \left( \beta^2 - \beta_H^2 \right) \right]
          \nonumber \\
  &\simeq& - \ \frac{2^{\frac{5}{2}} v_9}{\beta_H^{10}}
    \int_0^{\Lambda} d \tau_2 \ \tau_2^{\frac{7}{2}}
      \nonumber \\
  &=& - \ \frac{2^{\frac{7}{2}} v_9 \Lambda^{\frac{9}{2}}}
    {3^2 \beta_H^{10}}.
\EA
This free energy gives the leading contribution to the finite temperature effective potential at the closed string vacuum $V_{c,eff}$.
\BE
  V_{c,eff} (\beta)
    \simeq - \ \frac{2^{\frac{7}{2}} v_9 \Lambda^{\frac{9}{2}}}
      {3^2 \beta_H^{10}}.
\label{eq:Vceff}
\EE
It is noteworthy that $V_{c,eff} (\beta)$ is finite in the $\beta \rightarrow \beta_H$ limit, since $\Lambda$ is a finite constant. If the Hagedorn transition is the first order phase transition, then the potential energy at the minimum of the winding tachyon potential is lower than this one.

Let us next consider the potential energy at the open string vacuum ${\bf T} = {\bf 0}$ ($T = 0$). The free energy for a single D9-$\overline{\textrm{D9}}$ pair near the Hagedorn temperature is approximated as (\ref{eq:FHag}), since small $\tau$ part gives the leading term. Substituting $T=0$, we obtain
\BA
  F_o \left( T = 0 , \beta \right)
    &\simeq& - \ \frac{4 v_9}
      {\beta_H^{10}} \int_{0}^{\infty} dt \ 
        \exp \left( - \pi \ 
          \frac{\beta^2 - \beta_H^2}
            {\beta_H^2} \ t \right) \\
  &\simeq& - \ \frac{2 v_9}
    {\pi \beta_H^9
      \left( \beta - \beta_H \right)}.
\EA
The finite temperature effective potential is given by
\BA
  V_{o,eff} \simeq V (T) + F_o (T , \beta),
\EA
up to one-loop order, and it is approximated as
\BA
  V_{o,eff} \left( T = 0 , \beta \right)
    \simeq 2 \tau_9 v_9 
      - \ \frac{2 v_9}
        {\pi \beta_H^9
          \left( \beta - \beta_H \right)}.
\EA
In the case of $N$ D9-$\overline{\textrm{D9}}$ pairs, we can compute the finite temperature effective potential similarly \cite{Hotta6}. In this case tachyon field ${\bf T}$ is an $N \times N$ complex matrix. At ${\bf T} = {\bf 0}$, the potential energy is given by\footnote{The one-loop free energy of closed string gas might contribute this potential. However, the contribution is very small compared to that of open string gas.}
\BE
  V_{o,eff} \left( {\bf T} = {\bf 0} , \beta \right)
    \simeq 2 N \tau_9 v_9
      - \ \frac{2 N^2 v_9}
        {\pi \beta_H^9
          \left( \beta - \beta_H \right)}.
\label{eq:Voeff}
\EE
From this we can see that this potential energy decreases limitlessly as the temperature approaches to the Hagedorn temperature. Therefore, the potential energy at the open string vacuum becomes lower than that at the closed string vacuum (\ref{eq:Vceff}). It is preferable that we compare this potential energy with that at the minimum of the winding tachyon potential as we have mentioned before. However, the result indicates that, at sufficiently close to the Hagedorn temperature, the potential energy at the open string vacuum becomes lower than that at the minimum of the winding tachyon potential as long as the latter is finite. It is natural to think that the open string vacuum becomes the global minimum near the Hagedorn temperature. It is important to note that this calculation does not depend on the choice of Weyl factor, since the boundary term vanishes at ${\bf T} = {\bf 0}$.

\section{Conclusions and Discussions}
\label{sec:conclusion}

In this paper we have discussed the relation between the Hagedorn transition of closed strings and the thermal brane creation transition. The main subject of this paper is to propose the conjecture that {\it D9-brane--$\overline{\textrm{D9}}$-brane pairs are created by the Hagedorn transition of closed strings.} In other words, the open string vacuum becomes the stable minimum of the Hagedorn transition near the Hagedorn temperature.

Then we have shown some circumstantial evidences for this conjecture. First, we have shown that, in the Matsubara formalism, two types of amplitude of open strings approaches to closed string ones if we take an appropriate limit. The one-loop free energy of open strings close to the closed string vacuum has the form of the propagator of winding tachyon. The sphere amplitude for two winding tachyons vanishes, and the cylinder amplitude also vanishes if we adopt (\ref{eq:limit1}) in the closed string vacuum limit $|T| \rightarrow \infty$ together with the Hagedorn temperature limit $\beta \rightarrow \beta_H$. If we take these two limits under the condition (\ref{eq:limit2}), the cylinder amplitude with a single massless boson insertion approaches to the sphere amplitude with two winding tachyons and a single massless boson insertion. These are examples that we can identify the open string amplitude in the closed string vacuum limit with the closed string sphere amplitude with some winding tachyons insertion. It seems reasonable to conclude that we can identify the winding tachyon as the closed string vacuum limit of the boundary of an open string, which winds once around the compactified Euclidean time, though we need to investigate much other type of amplitudes in order to confirm it.

Secondly, we have evaluated the potential energy at the closed string vacuum and at the open string vacuum. We have shown that the potential energy at the open string vacuum decreases limitlessly as the temperature approaches to the Hagedorn temperature. Therefore, the potential energy at the open string vacuum becomes lower than that at the closed string vacuum as long as the latter is finite. From this we may say that the open string vacuum becomes the global minimum in entire space of the open string tachyon field near the Hagedorn temperature. This is the property that the stable minimum of the Hagedorn transition is expected to have.

We have discussed only type IIB string theory case in this paper. However, we expect that almost the same argument holds for the type IIA string theory case. That is to say, non-BPS D9-branes are created by the Hagedorn transition of closed strings. We need to check this case precisely including the overall numerical factors of amplitudes and so on.

In the calculation of the one-loop free energy of open strings in the framework of BSFT, we need to overcome the problem of the choice of the Weyl factors in the two boundaries of the one-loop world-sheet \cite{1loopAO} \cite{1loopann1} \cite{1loopann2} \cite{1loopsym1} \cite{1loopsym2} \cite{1loop1} \cite{1loop2} \cite{1loop3} \cite{1loop4}. However, if we can reproduce the same result by using another method, it reinforces our previous studies. We have computed free energy of open strings on a D-brane--anti-D-brane pair based on thermo field dynamics (TFD). This free energy agrees with that based on Matsubara formalism \cite{Hotta9}. In the framework of TFD, we can derive free energy in the ideal gas approximation without handling with loop calculation.

We can also analyze D-brane--anti-D-brane pairs at finite temperature by using boundary conformal field theory (BCFT). We are not confronted with a problem of the choice of the Weyl factors if we compute amplitudes in the framework of BCFT. The thermodynamic properties are investigated by using BCFT as S-brane thermodynamics \cite{SbraneThermo1} \cite{SbraneThermo2}. In this case the boundary action is chosen as that corresponding to the Wick rotated version of rolling tachyon configuration. Previously, the one-loop partition function is calculated based on BCFT by Sugawara \cite{oneloopSbrane}. Furthermore, Gaiotto, Itzhaki and Rastelli have advocated that the disk scattering amplitude of $m$ closed string at the closed string vacuum is equivalent to a sphere amplitude with $m+1$ closed string insertion in the context of imaginary D-branes \cite{ImagD}. Their calculation might be useful if we compute various amplitudes in the Matsubara formalism on the basis of BCFT, instead of BSFT.

As we have mentioned in \S \ref{sec:DDbar}, we have investigated the thermal brane creation transition in the case of multiple D9-$\overline{\textrm{D9}}$ pairs previously. We have concluded that a large number $N$ of D9-$\overline{\textrm{D9}}$ pairs are created simultaneously, since the critical temperature is a decreasing function of $N$ as long as the 't Hooft coupling is very small. However, this means that we have to perform a non-perturbative calculation in order to investigate the thermodynamical behavior of strings and branes in the final stage of the phase transition. We need to evaluate the potential energy at the open string vacuum non-perturbatively. It would be interesting to study based on, for example, the matrix model \cite{matrix}, the IIB matrix model \cite{IIBmatrix} and the USp matrix model \cite{USp1} \cite{USp2}. The K-matrix model might also be useful, as it explicitly contains open string tachyon \cite{Kmatrix1} \cite{Kmatrix2} \cite{Kmatrix3}.

The correspondence of amplitude which we have calculated in \S \ref{sec:ClosedStringVacuum} is satisfied only in the case of a single D9-$\overline{\textrm{D9}}$ pair, instead of multiple D9-$\overline{\textrm{D9}}$ pairs. We need to check whether this result is consistent with the previous result that multiple D9-$\overline{\textrm{D9}}$ pairs are created simultaneously in the thermal brane creation transition. For this reason, we have to investigate thermodynamic properties of strings close to closed string vacuum more precisely.

Finally, the phase transition from closed strings to spacetime-filling branes is reminiscent of the phase transition in the Plank solid model of Schwarzschild black holes \cite{Hotta3}. In order to show the validity of this model, we need to show that closed strings are excited enough to reach the critical temperature of the Hagedorn transition in the vicinity of the would-be horizon, and that the world volume theory of D9-$\overline{\textrm{D9}}$ pairs at large 't Hooft coupling at high temperature is effectively described by topological field theories. Originally Witten has constructed topological field theories in order to describe unbroken phase of quantum gravity or string theory \cite{TFT} \cite{TSM} \cite{TG1} \cite{TG2} \cite{TG3}. In this phase it is expected that general covariance is confined and unbroken, and there are no propagating degrees of freedom. In this sense, we would like to show that the unbroken phase of strings consists of a large number $N$ of D9-$\overline{\textrm{D9}}$ pairs. It would be also interesting to investigate the relation between our work and phase of ``nothing", which is expected to be generated via the condensation of closed string winding tachyon \cite{phaseofnothing1} \cite{phaseofnothing2}.

\section*{Acknowledgements}

The author would like to thank K. Ishikawa and H. Itoyama for useful discussions and encouragement. He also thanks colleagues at Hokkaido University for useful discussions. He appreciates the Yukawa Institute for Theoretical Physics at Kyoto University. Discussions during the YITP workshop YITP-T-18-04 on "New Frontiers in String Theory 2018" were useful to complete this work.

\appendix

\section{S-representation and F-representation}
\label{sec:SF}

In this appendix we briefly explain the transformation between one-loop free energy of closed string gas in the S-representation and that in the F-representation by using unfolding technique \cite{Tan1} \cite{McRoth}. In the bosonic string and the heterotic string cases, the transformation is shown in Ref. \cite{Tan1}. We see how this unfolding technique are applied to the transformation in our type II superstring case in this appendix.

The transformation is shown by using PSL(2,Z) modular transformation. This transformation is generated by shift $T$ and inversion $S$, which is given by (\ref{eq:modular1}) and (\ref{eq:modular2}), respectively. A general transformation is given by
\BA
  \tau' = M (\tau) = \frac{a \tau + b}{c \tau + d},
\label{eq:Mtau}
\EA
where $a$, $b$, $c$ and $d$ are integers, and satisfy
\BA
  ad - bc = 1.
\label{eq:SL2Zcond}
\EA
This can be represented $2 \times 2$ matrix as
\BA
  M &=&
    \left(
      \begin{array}{cc}
        a & b \\
        c & d \\
      \end{array}
    \right).
\EA
Let us denote the greatest common divisor of integers $x$ and $y$ as $[x,y]$ as we have done in \S \ref{sec:HagTra}. From the condition (\ref{eq:SL2Zcond}), we can easily show that the integers $c$ and $d$ satisfy
\BA
  [c,d] = 1,
\label{eq:cdrelprime}
\EA
that is, $c$ and $d$ are relatively prime. Two transformations $M$ and $M'$, which have the same lower row entries $(c,d)$, are related by the shift $T$. This can be shown as follows. If we denote the upper row entries of $M$ and $M'$ as $(a,b)$ and $(a',b')$, respectively, then the condition (\ref{eq:SL2Zcond}) for $M$ and $M'$ are written as $ad - bc = 1$ and $a'd - b'c = 1$. These integers satisfy $(a - a') d = (b - b') c$. Since $c$ and $d$ are relatively prime, $a$ and $b$ are represented as $a = a' + lc$ and $b = b' + ld$, respectively, where $l$ is an appropriate integer. This is exactly the succession of the shift transformations $l$ times, i.e., $T^l$. The modular transformations are classified into families labeled by $c$ and $d$. We denote the transformation $M$ whose lower row entries are $(c,d)$ as $M (c,d)$.

In order to show the relation between F-representation and S-representation, we need to transform the domain of integration from fundamental region ${\cal F}$ into half strip region ${\cal S}$ (see Figure \ref{fig:Srep} and Figure \ref{fig:Frep}). Within each family, there exists a unique transformation $Q (c , d) \in M (c,d)$ which maps the fundamental domain ${\cal F}$ into the half strip region ${\cal S}$. Let us denote the image of ${\cal F}$ under $Q (c , d)$ as ${\cal S} (c,d)$. The union of ${\cal S} (c,d)$ makes up the entire half strip region \cite{Tan1}, namely,
\BA
  {\cal S} = \bigcup_{c,d} {\cal S} (c,d).
\EA
Thus, if we can show that the integrand in (\ref{eq:closedFrep}) is transformed to that in (\ref{eq:closedSrep}) by $Q (c , d)$, this transformation connects the F-representation and S-representation.

Now, we describe the relation between the sum over $m$ and $n$ in the F-representation (\ref{eq:closedFrep}) and the sum over $w$ in the S-representation (\ref{eq:closedSrep}). These integers are connected by $Q (c , d)$. From (\ref{eq:Mtau}), the transformation of $\tau_2$ reads
\BA
  \tau_2 ' = \frac{\tau_2}{\left( c \tau_1 + d \right)^2 + (c \tau_2)^2}.
\EA
It should be noted that this transformation does not depend on $a$ and $b$. Applying this transformation to the sum over $w$ (without $(-1)^w$) in the one-loop free energy in the S-representation (\ref{eq:closedSrep}), we obtain
\BA
  \sum_{w=1}^{\infty}
    \exp \left(- \ \frac{2 \pi w^2 \beta^2}
      {\beta_H^2 \tau_2} \right)
        &=& \sum_{w=1}^{\infty}
          \exp \left[ - \ \frac{2 \pi w^2 \beta^2}
            {\beta_H^2 \tau_2}
              \left\{ \left( c \tau_1 + d \right)^2 + (c \tau_2)^2
                \right\} \right]
                  \nonumber \\
  &=& \sum_{w=1}^{\infty}
    \exp \left[ - \ \frac{2 \pi \beta^2}
      {\beta_H^2 \tau_2}
        \left\{ \left( wc \tau_1 + wd \right)^2 + (wc \tau_2)^2
          \right\} \right],
\EA
where we have assumed that the factors depending on $c$ or $d$ do not arise from the other factors in (\ref{eq:closedSrep}). We here introduce the relation between $m$, $n$, $w$, $c$ and $d$ as \cite{Tan1}
\BA
  [m,n] &=& w,
\label{eq:mnw} \\
  m &=& wc,
\label{eq:mwc} \\
  n &=& wd.
\label{eq:nwd}
\EA
We can derive the sum over $m$ and $n$ in the one-loop free energy in the F-representation (\ref{eq:closedFrep}) by taking sum not only over $w$ but also over relatively prime pair of integers $(c,d)$:
\BA
  \sum_{c , \ d , \ [c,d] = 1} \ \sum_{w=1}^{\infty}
    \exp \left(- \ \frac{2 \pi w^2 \beta^2}{\beta_H^2 \tau_2} \right)
      &=& {\sum_{m,n = - \infty}^{\infty}}'
        \exp \left[ - \ \frac{2 \pi \beta^2}{\beta_H^2 \tau_2}
          \left\{ \left( m \tau_1 + n \right)^2 + (m \tau_2)^2
            \right\} \right]
              \nonumber \\
  &=& {\sum_{m,n = - \infty}^{\infty}}'
    \exp \left[ - \ \frac{2 \pi \beta^2}{\beta_H^2 \tau_2}
      \left( n^2 + m^2 |\tau| + 2 \tau_1 mn \right) \right]
        \nonumber \\
  &=& {\sum_{m,n = - \infty}^{\infty}}' e^{- S_{\beta} (m,n)}.
\EA
This factor corresponds to the first term in the braces in (\ref{eq:closedFrep}). The sum over $c$ and $d$ comes from the union of the domain of integration ${\cal S} (c,d)$. We can derive the other terms in the braces in (\ref{eq:closedFrep}) from (\ref{eq:closedSrep}) by performing similar calculation.

For concreteness, let us consider the case that $(c,d) = (-1,0)$ as an example. In this case the modular transformation $Q (-1,0)$ is the inversion $S$, namely,
\BA
  Q (-1,0) : \ \ \ \tau \rightarrow \tau ' = - \ \frac{1}{\tau}.
\label{eq:Q(-1,0)}
\EA
Let us denote the one-loop free energy in the S-representation (\ref{eq:closedSrep}), whose domain of integration is restricted to ${\cal S} (c,d)$, as $F_{{\cal S} (c,d)}$. In our case, the domain of integration ${\cal S} (-1,0)$ is depicted in Figure \ref{fig:S(-1,0)}. Applying (\ref{eq:Q(-1,0)}) to $F_{{\cal S} (-1,0)}$, we obtain
\begin{figure}[tbp]
\centering
\includegraphics[width=.5\textwidth]{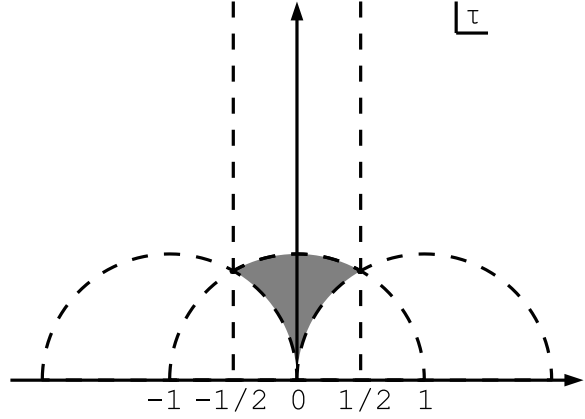}
\caption{\label{fig:S(-1,0)} Shaded region is the domain of integration ${\cal S} (-1,0)$.}
\end{figure}
\BA
  F_{{\cal S} (-1,0)} (\beta)
    &=& - \ \frac{8 (2 \pi)^8 v_9}{\beta_H^{10}}
      \int_{{\cal S} (-1,0)} \frac{d^2 \tau '}{{\tau_2 '}^2} \ 
        \frac{1}{{\tau_2 '}^4} \ 
          \frac{1}{\left| {\vartheta_1}' (0 | \tau ') \right|^8}
            \nonumber \\
  && \hspace{3mm}
    \times \left[ \left\{ \left( \vartheta_3^4 - \vartheta_4^4 \right)
      \left( {\bar{\vartheta}}_3^4 - {\bar{\vartheta}}_4^4 \right)
        + \vartheta_2^4 {\bar{\vartheta}}_2^4 \right\} (0 | \tau ') \ 
          \sum_{w=1}^{\infty}
            \exp \left( - \ \frac{2 \pi w^2 \beta^2}
              {\beta_H^2 \tau_2 '} \right)
                \right. \nonumber \\
  && \hspace{3mm} \left.
    - \left\{ \left( \vartheta_3^4 - \vartheta_4^4 \right)
      {\bar{\vartheta}}_2^4 + \vartheta_2^4
        \left( {\bar{\vartheta}}_3^4
          - {\bar{\vartheta}}_4^4 \right) \right\} (0 | \tau ') \ 
            \sum_{w=1}^{\infty} (-1)^w
              \exp \left( - \ \frac{2 \pi w^2 \beta^2}
                {\beta_H^2 \tau_2 '} \right) \right]
                  \nonumber \\
  &=& - \ \frac{8 (2 \pi)^8 v_9}{\beta_H^{10}}
    \int_{{\cal F}} \frac{d^2 \tau}{{\tau_2}^2} \ 
      \frac{(\tau_1^2 + \tau_2^2)^4}{{\tau_2}^4} \ 
        \frac{1}{\left| {\vartheta_1}'
          \left( 0 \left| - \ \frac{1}{\tau} \right. \right) \right|^8}
            \nonumber \\
  && \hspace{30mm}
    \times \left[ \left\{ \left( \vartheta_3^4 - \vartheta_4^4 \right)
      \left( {\bar{\vartheta}}_3^4 - {\bar{\vartheta}}_4^4 \right)
        + \vartheta_2^4 {\bar{\vartheta}}_2^4 \right\}
          \left( 0 \left| - \ \frac{1}{\tau} \right. \right)
            \right. \nonumber \\
  && \hspace{60mm}
    \times \sum_{w=1}^{\infty}
      \exp \left( - \ \frac{2 \pi w^2 \beta^2}{\beta_H^2 \tau_2}
        (\tau_1^2 + \tau_2^2) \right)
          \nonumber \\
  && \hspace{30mm}
    - \left\{ \left( \vartheta_3^4 - \vartheta_4^4 \right)
      {\bar{\vartheta}}_2^4 + \vartheta_2^4
        \left( {\bar{\vartheta}}_3^4
          - {\bar{\vartheta}}_4^4 \right) \right\}
            \left( 0 \left| - \ \frac{1}{\tau} \right. \right)
              \nonumber \\
  && \hspace{60mm} \left.
    \times \sum_{w=1}^{\infty} (-1)^w
      \exp \left(- \ \frac{2 \pi w^2 \beta^2}{\beta_H^2 \tau_2}
        (\tau_1^2 + \tau_2^2) \right) \right]
          \nonumber \\
  &=& - \ \frac{8 (2 \pi)^8 v_9}{\beta_H^{10}}
    \int_{{\cal F}} \frac{d^2 \tau}{{\tau_2}^6} \ 
      \frac{1}{\left| {\vartheta_1}' (0 | \tau) \right|^8}
        \nonumber \\
  && \hspace{20mm}
    \times \left[ \left\{ \left( \vartheta_3^4 - \vartheta_2^4 \right)
      \left( {\bar{\vartheta}}_3^4 - {\bar{\vartheta}}_2^4 \right)
        + \vartheta_4^4 {\bar{\vartheta}}_4^4 \right\} (0 | \tau)
          \sum_{w=1}^{\infty} e^{- S_{\beta} (-w , 0)}
            \right. \nonumber \\
  && \hspace{20mm} \left.
    - \left\{ \left( \vartheta_3^4 - \vartheta_2^4 \right)
      {\bar{\vartheta}}_4^4 + \vartheta_4^4
        \left( {\bar{\vartheta}}_3^4
          - {\bar{\vartheta}}_2^4 \right) \right\} (0 | \tau)
            \sum_{w=1}^{\infty} (-1)^w e^{- S_{\beta} (-w , 0)}
              \right],
                \nonumber \\
\EA
where we have used the modular transformation of the Jacobi theta functions (\ref{eq:thetaS1}) $\sim$ (\ref{eq:thetaS4}), and the invariance of $d^2 \tau / {\tau_2}^2$ under PSL(2,Z) transformation. On the other hand, if we denote the one-loop free energy in the F-representation (\ref{eq:closedFrep}) restricted to one pair of (c,d) using the relation (\ref{eq:mwc}) and (\ref{eq:nwd}) as $F_{{\cal F} (c,d)}$, then $F_{{\cal F} (-1,0)}$ is represented as
\BA
  F_{{\cal F} (-1,0)} (\beta)
    &=& - \ \frac{8 (2 \pi)^8 v_9}{\beta_H^{10}}
      \int_{\cal F} \frac{d^2 \tau}{\tau_2^6}
        \frac{1}{\left| {\vartheta_1}' (0 | \tau) \right|^8} \ 
          {\sum_{w=1}^{\infty}}' e^{- S_{\beta} (-w,0)}
            \nonumber \\
  && \times \left\{ \left( \vartheta_2^4 {\bar{\vartheta}}_2^4
    + \vartheta_3^4 {\bar{\vartheta}}_3^4
      + \vartheta_4^4 {\bar{\vartheta}}_4^4 \right) (0 | \tau)
        + (-1)^{-w} \left( \vartheta_2^4 {\bar{\vartheta}}_4^4
          + \vartheta_4^4 {\bar{\vartheta}}_2^4 \right) (0 | \tau)
            \right. \nonumber \\
  && \hspace{10mm} \left.
    - (-1)^{-w} \left( \vartheta_2^4 {\bar{\vartheta}}_3^4
      + \vartheta_3^4 {\bar{\vartheta}}_2^4 \right) (0 | \tau)
        - \left( \vartheta_3^4 {\bar{\vartheta}}_4^4
          + \vartheta_4^4 {\bar{\vartheta}}_3^4 \right) (0 | \tau)
            \right\}.
\EA
From this we can see that $F_{{\cal S} (-1,0)}$ agrees with $F_{{\cal F} (-1,0)}$. Similarly, we can show that the same relation between $F_{{\cal S} (c,d)}$ and $F_{{\cal S} (c,d)}$, that is,
\BA
  F_{{\cal S} (c,d)} = F_{{\cal F} (c,d)}.
\EA
The one-loop free energy in the S-representation is the sum of $F_{{\cal S} (c,d)}$ over relatively prime pair of integers $(c,d)$, and that in the F-representation is also the sum of $F_{{\cal F} (c,d)}$. We therefore arrive at the conclusion that the one-loop free energy in the S-representation (\ref{eq:closedSrep}) agrees with that in the F-representation (\ref{eq:closedFrep}). It should be noted that we can not specify what mode in the S-representation corresponds to each $(m,n)$-mode in the F-representation. Each $(m,n)$-mode corresponds to $w$-mode whose domain of integration is restricted to the union of ${\cal S} (c,d)$, where $c$, $d$ and $w$ satisfy the relation (\ref{eq:mnw}) $\sim$ (\ref{eq:nwd}).

Finally, we explain the construction of $Q(c,d)$ for each $c$ and $d$ from the series of the shift $T$ and the inversion $S$ \cite{Tan1}. Let us consider $Q (-c , d)$ in the case that $d > c > 1$, for example. We choose $-c$ instead of $c$ for the later convenience. In order to see the construction of $Q(-c,d)$, it is useful to prepare a finite sequence of sets of positive integers $(n_1 ; c_1 , d_1)$, $(n_2 ; c_2 , d_2)$, $\cdots$, $(n_f ; c_f , d_f)$. From $c$ and $d$, we define the first set $(n_1 ; c_1 , d_1)$ as
\BA
  n_1 &=& \left[ \frac{d}{c} \right], \\
  c_1 &=& n_1 c - d, \\
  d_1 &=& c,
\EA
where $[x]$ denotes the smallest integer $n$ with $n \geq x$. By using Euclidean algorithm, we obtain
\BA
  [c_1 , d_1] = [n_1 c - d , c] = [-d , c] = 1.
\EA
Since $c \neq 1$, $c > c_1 \geq 1$. Similarly, we can define recursively $(n_i ; c_i , d_i)$ as
\BA
  n_i &=& \left[ \frac{d_{i-1}}{c_{i-1}} \right], \\
  c_i &=& n_i c_{i-1} - d_{i-1}, \\
  d_i &=& c_{i-1}.
\EA
If $c_{i-1} \neq 1$, then $c_{i-1} > c_i > 0$. Since
\BA
  [c_i , d_i] = [n_i c_{i-1} - d_{i-1} , c_{i-1}] = [-d_{i-1} , c_{i-1}] = 1,
\EA
$c_i$ and $d_i$ is relatively prime. When $c_{i-1} = 1$, we denote such an index $i$ as $f$, and $n_f = d_{f-1}$, $c_f = 0$, $d_f = 1$. We obtain the series of integers $(n_i ; c_i , d_i)$ like this. By using this sequence of positive integers, we can represent $Q (-c,d)$ as
\BA
  Q (-c , d) = T^j M_f (- c_f, d_f) M_{f-1} (- c_{f-1}, d_{f-1})
    \cdots M_2 (- c_2, d_2) M_1 (- c_1, d_1),
\EA
where $j$ is an appropriate integer. We need $T^j$ in order to guarantee that this transformation maps the fundamental domain ${\cal F}$ into the half strip region ${\cal S}$. For the case other than $d > c \geq 1$, we can boil down to this case by operating $K$ at first, where $K$ is a product of $-I$, $TS$ or $(TS)^{-1}$ \cite{Tan1}. Thus, $Q (-c,d)$ can be represented as
\BA
  Q (-c , d) = T^j M_f (- c_f, d_f) M_{f-1} (- c_{f-1}, d_{f-1})
    \cdots M_2 (- c_2, d_2) M_1 (- c_1, d_1) K.
\EA
In this way, $Q (-c,d)$ are concretely constructed from the generators of PSL(2,Z) modular transformation for each relatively prime pair of integers $(c,d)$. By using these $Q (-c,d)$'s, we can transform the one-loop free energy of closed string gas in the S-representation (\ref{eq:closedSrep}) into that in the F-representation (\ref{eq:closedFrep}).

\acknowledgments

The author would like to thank H. Itoyama and M. Kato for useful discussions and encouragement. He also thanks colleagues at Hokkaido University for useful discussions. He appreciates the Yukawa Institute for Theoretical Physics at Kyoto University. Discussions during the YITP workshop YITP-T-18-04 on "New Frontiers in String Theory 2018" were useful to complete this work.

\end{document}